\documentclass[11pt,letter]{article}
\usepackage{graphicx}
\usepackage{epstopdf}
\usepackage{epsfig,amssymb}
\usepackage{xcolor}
\usepackage[cm]{fullpage}

\definecolor{plum}{rgb}{0.36078, 0.20784, 0.4}
\definecolor{chameleon}{rgb}{0.30588, 0.60392, 0.023529}
\definecolor{cornflower}{rgb}{0.12549, 0.29020, 0.52941}
\definecolor{scarlet}{rgb}{0.8, 0, 0}
\definecolor{brick}{rgb}{0.64314, 0, 0}
\newcommand{\email}[1]{\href{mailto:#1}{\tt \textcolor{cornflower}{#1}}}

\usepackage[%
    pdftitle={On the Stress Tensor for Asymptotically Flat Gravity},%
    pdfauthor={Robert Mann, Donald Marolf, Robert McNees, Amitabh Virmani},%
    pdfsubject={Document Subject},%
    pdfkeywords={Document Keywords},
    colorlinks=true,linkcolor=cornflower,citecolor=scarlet,urlcolor=chameleon%
]{hyperref}

\newcommand{\beq}{\begin{equation}}
\newcommand{\eeq}{\end{equation}}
\newcommand{\be}{\begin{equation}}
\newcommand{\ee}{\end{equation}}
\newcommand{\beqa}{\begin{eqnarray}}
\newcommand{\eeqa}{\end{eqnarray}}
\newcommand{\beqar}{\begin{eqnarray*}}
\newcommand{\eeqar}{\end{eqnarray*}}
\newcommand{\bea}{\begin{eqnarray}}
\newcommand{\eea}{\end{eqnarray}}

\newcommand{\s}{\sigma}

\newcommand{\nn}{\nonumber}
\newcommand\nts{\!\!}

\newcommand\defeq{\mathrel{\mathop:}=}

\begin{document}

\setlength{\unitlength}{1mm}

\begin{titlepage}

\begin{flushright}
\today
\end{flushright}
~\vspace{3cm}\\

\begin{center}
{\bf \Large On the Stress Tensor for Asymptotically Flat Gravity}
\end{center}

\vspace{1cm}

\begin{center}
Robert B. Mann $^{a, b}$, Donald Marolf $^{c}$, Robert McNees $^{b}$, and Amitabh
Virmani $^{c}$

\vspace{1cm}{\small {\textit{$^{a}$Dept. of Physics, University of
Waterloo, Waterloo, Ontario N2L 3G1, Canada}}}\\
\vspace{2mm} {\small {\textit{$^{b}$Perimeter Institute for Theoretical Physics, 31 Caroline Street North, Waterloo,
Ontario N2L 2Y5, Canada}}}\\
\vspace{2mm} {\small {\textit{$^{c}$Department of Physics,
University of California, Santa Barbara, CA 93106-9530,
USA}}}\\
\vspace*{0.5cm}
\email{rbmann@sciborg.uwaterloo.ca}\,,
\email{marolf@physics.ucsb.edu}\,,\\
\email{rmcnees@perimeterinstitute.ca}\,,
\email{virmani@physics.ucsb.edu}
\end{center}
\vspace{0.5cm}

\begin{abstract}
The recent introduction of a boundary stress tensor for
asymptotically flat spacetimes enabled a new construction of energy,
momentum, and Lorentz charges. These charges are known to generate
the asymptotic symmetries of the theory, but their explicit formulas
are not identical to previous constructions in the literature. This paper
corrects an earlier comparison with other approaches, including terms in the definition of the stress tensor charges that were previously overlooked. We show that
these terms either vanish identically (for $d > 4$) or take a form that does not contribute to the conserved charges (for $d=4$). This verifies the earlier claim that boundary
stress tensor methods for asymptotically flat spacetimes yield the
same conserved charges as other approaches.  We also derive some
additional connections between the boundary stress tensor and the
electric part of the Weyl tensor.
\end{abstract}

\end{titlepage}

\vspace{0.5cm} \tableofcontents


\setcounter{equation}{0}


\section{Introduction}
\label{introduction}

Conserved quantities play an important role in our description of
physical systems.  Particularly interesting cases arise in gauge
theories, where conserved charges may be expressed as surface
integrals at infinity. For a theory that is diffeomorphism invariant, like
general relativity, the construction of conserved quantities
depends crucially  on the asymptotic structure of spacetime.
In the case of asymptotically flat spacetimes in $d \geq 4$ dimensions, one is most interested in conserved charges associated with the generators of asymptotic Poincar\'e transformations.

There are many different methods for constructing conserved charges in asymptotically flat gravity \cite{ADM1,RT,AH,A,AR,AD,Sorkin,HH,Mann,MM}.  The various approaches offer different perspectives, but ultimately lead to equivalent results.
For example, Arnowitt, Deser, and Misner (ADM) \cite{ADM1,RT} based their construction on the
initial value formulation of general relativity. This introduced a notion of asymptotic flatness at spatial infinity by
demanding that the Cauchy data on some initial surface (rather than
the geometry of the full spacetime) should asymptotically approach
that of a corresponding surface in Minkowski space.  In contrast, a
fully covariant approach to asymptotically flat conserved quantities
was introduced by Ashtekar and Hansen  in the late 1970s. Several
important results were established within this framework, including
the introduction of a precise notion of asymptotic flatness at
spatial infinity \cite{AH,A}, and showing that the ADM 4-momentum is
obtained from the past limit of Bondi 4-momentum \cite{AM1}. More recently, a definition of conserved charges was introduced that makes use of a boundary stress tensor \cite{MM}.  Boundary stress tensor methods for asymptotically flat space were motivated by the success of such
methods in asymptotically anti-de Sitter space (see \cite{skenderis,Kraus,Emparan:1999pm} as well as \cite{Mann} and many succeeding works), which were in turn inspired by the methods of \cite{BY}. Earlier steps toward
using such methods in asymptotically flat space were taken in
\cite{Mann,KLS,skenAF,Mann2}.

The starting point of the boundary stress tensor method is a valid variational principle for asymptotically flat gravity. The standard form of the action, comprising the Einstein-Hilbert and Gibbons-Hawking-York terms, must be modified because it is not stationary with respect to arbitrary field variations that preserve the boundary conditions. This is directly related to the problem investigated by Regge and Teitelboim in \cite{RT}, who found that the Hamiltonian formulation of general relativity with asymptotically flat boundary conditions is only well-defined if it includes the additional surface term first proposed in \cite{ADM1}. The same conclusion applies to the Lagrangian formulation of the theory, but the necessary boundary terms were not investigated until recently; see  \cite{MM} for a discussion in the 2nd order formalism or \cite{eeA,ABL,AshSloan} for a discussion in the first order formalism. Given a valid variational principle, the boundary stress tensor $T_{ab}$ is defined by varying the on-shell action with respect to the boundary metric, as in \cite{BY,skenderis,Kraus}. Any asymptotic Killing field $\xi^{a}$ can then be used to define a current $T_{ab}\,\xi^{b}$, and the flux of this current across a cut $C$ of spatial infinity is a conserved charge. According to the results of \cite{HIM} these charges necessarily generate the asymptotic symmetries of the theory. In \cite{MM} it was argued -- on general grounds -- that the conserved quantities defined in this way must agree with other definitions. To illustrate this point the boundary stress tensor definition of energy and momentum at spatial infinity were shown to agree with the charges defined by Ashtekar and Hansen in \cite{AH}.

The comparison between the boundary stress tensor approach of \cite{MM} and other methods in the literature was extended in \cite{MMV}, where two important results were established. First, it was shown that the canonical (space + time) reduction of the covariant action yields the ADM result \cite{ADM1} with the appropriate boundary terms.  Second, using the form
\be \label{InitialStressTensor}
T_{ab}=\frac{1}{8\pi G} \left( \pi_{ab} - \hat{\pi}_{ab}\right)
\ee
of the  boundary stress tensor presented in \cite{MM}, explicit formulas for boundary stress tensor charges associated with asymptotic translations and Lorentz transformations  were shown to agree with those given  by  Ashtekar and Hansen in \cite{AH}. In (\ref{InitialStressTensor}) $\pi_{ab}$ is the conjugate momentum of the gravitational field and $\hat{\pi}_{ab}$ is an analogous contribution from the counterterms. However, it turns out that there are several additional terms (beyond those in (\ref{InitialStressTensor})) which potentially contribute to $T_{ab}$ but were overlooked in \cite{MM}\,\footnote{We thank Julian Le Witt and Simon Ross for bringing this fact to our
attention.}.

The primary purpose of this paper is to address the terms in the stress tensor that were overlooked in \cite{MM,MMV}. Specifically, we want to resolve the tension between these `extra terms' and the observation in \cite{MMV} that the stress tensor (\ref{InitialStressTensor}) leads to charges that agree with the definition of Ashtekar and Hansen \cite{AH}. To this end, we explicitly calculate the extra terms and show that they do not contribute to the conserved charges. In $d>4$ spacetime dimensions this is because the extra terms are identically zero.  However, for $d=4$ the situation is more subtle; the extra terms are not zero, but they do not contribute to the conserved charges.
We also establish several additional results that we summarize here in simple physical terms.
\begin{itemize}
\item For $d>4$ the boundary stress tensor agrees with the electric part of the Weyl tensor at both first and second order in an asymptotic expansion.
 \item For $d=4$ the first order term in the asymptotic expansions of the boundary stress tensor and the electric part of the Weyl tensor are the same.  This result also applies to spacetimes carrying non-zero NUT charge. At second order the boundary stress tensor is given by the second order term in the electric part of the Weyl tensor, plus contributions from the extra terms described above. The extra terms vanish for some spacetimes and are non-zero for others.  However, in no case do they contribute to the conserved quantities.
\item For $d=4$, with zero NUT charge, there is a non-trivial identity relating terms in the asymptotic expansions of the electric and magnetic parts of the Weyl tensor
     \be
      E^{_{(2)}}_{ab} - \s E^{_{(1)}}_{ab}= \epsilon_{cd(a}D^c \beta_{b)}^{\,\,d}~.
     \ee
     This identity helps one see more clearly the equivalence between the boundary stress tensor and Ashtekar-Hansen charges. It can also be used to simplify the analysis of \cite{MMV}.
\end{itemize}

The rest of the paper is organized as follows.  We begin with
various definitions and a brief review of asymptotic flatness in
section \ref{prelims}.  We then present the covariant
counterterm action of \cite{MM} in section \ref{varprinc}, and vary it with respect to the boundary metric to obtain the full boundary stress tensor. In sections  \ref{bst4} and
\ref{bstd} we discuss the asymptotic expansion of the stress tensor
in $d=4$ and $d>4$ dimensions, respectively. These sections refer to calculations that are presented in detail in \ref{bstd=4} and \ref{astdg4}. The stress-tensors are then used to construct conserved quantities at spatial infinity that agree with the usual definitions \cite{ADM1, RT, AH, A, AR} of energy, momentum, angular momentum, and boosts. We establish this by
explicitly showing that the extra terms overlooked in \cite{MM, MMV}
do not contribute to the conserved charges. Finally we consider some
examples in section \ref{Examples} and close with a brief discussion
in section \ref{disc}. Throughout we use the notation of \cite{MM,
MMV, Wald}.

\setcounter{equation}{0}

\section{Preliminaries}
\label{prelims}

This section sets the stage for our later work by providing relevant
definitions and review.  In the first part we state our definition of
asymptotic flatness in four and higher dimensions, followed by
a discussion of the asymptotic field equations. The calculations in this part
follow the approach of Beig and Schmidt \cite{BS, B}. The second part contains
definitions and a few useful results concerning the electric part of the Weyl tensor.

\subsection{Asymptotic Flatness}

We begin by reviewing the coordinate-based definition of asymptotic flatness
that was used in \cite{MM}. This approach is inspired by the covariant phase space treatement of
\cite{ABR} and is modeled on the definitions given in \cite{BS,B}. Our definition is
particularly close to that of \cite{AR}, which treats spatial infinity as the unit time-like hyperboloid.

A spacetime of dimension $d \ge 4$ is asymptotically flat at spatial infinity if the line element admits an expansion of the form
\begin{equation}  \label{AFdef}
ds^2 = \left( 1+ \frac{2 \sigma}{\rho^{\,d-3}} + \mathcal{O}(\rho^{-(d-2)})%
\right) d\rho^{\,2} + \rho^{\,2} \left( h^{(0)}_{ab} + \frac{h^{(1)}_{ab}}{\rho^{d-3}}
+ \mathcal{O}(\rho^{-(d-2)}) \right) d\eta^a d\eta^b +  \rho \left( \mathcal{O}(\rho^{-(d-2)}) \right)_{a} d\rho \,d\eta^a,
\end{equation}
for large positive $\rho$, where $h^{_{(0)}}_{ab}$ and $\eta^a$ are a metric and the associated coordinates on the unit $(d-2,1)$ hyperboloid ${\cal H}$, and $\sigma, h^{_{(1)}}_{ab}$ are respectively a smooth function and a smooth tensor field on ${\cal H}$. The coordinate $\rho$ is the hyperbolic ``radial" function associated with some asymptotically Minkowski coordinates $x^{a}$ through $\rho^2 = \eta_{ab} x^a x^b.$ In (\ref{AFdef}), the symbols $\mathcal{O}(\rho^{-(d-2)})$ refer to terms that fall-off at least as fast as $\rho^{-(d-2)}$ as one approaches \textit{spacelike} infinity, i.e., $\rho \rightarrow +\infty$ with the coordinates $\eta^a$ fixed.

Following \cite{BS,B} we partially fix the gauge and bring the metric into the form
\begin{eqnarray}
\label{metric1}ds^2 &=& N^2 d\rho^2 +  h_{ab}d\eta^a d\eta^b
\\ &=&\label{metric2} \left( 1 + \frac{\s}{\rho^{\,d-3}} \right)^2 d\rho^{\,2} + \rho^{\,2} \left[
h^{(0)}_{ab} + \frac{h^{(1)}_{ab}}{\rho^{\,d-3}} +  \frac{h^{(2)}_{ab}}{\rho^{\,d-2}}
+ {\cal O}\left( \frac{1}{\rho^{\,d-1}}\right)  \right]d\eta^a d\eta^b,
\end{eqnarray} where again $h^{_{(0)}}_{ab}$ is the metric on the unit hyperboloid.
In four dimensions there are two additional conditions that must be imposed \cite{AH, B, ABR}.
First, we require $h^{_{(1)}}_{ab} = -2 \s h^{_{(0)}}_{ab}$. This rules out the
possibility of the so-called ``spi super-translations''
\cite{AH}. Second, the function $\s$ must be even under the natural
inversion mapping induced by the Minkowskian transformation $x^a
\rightarrow -x^{a}$, ruling out the so-called ``logarithmic
super-translations.'' These conditions ensure that the asymptotic
symmetry group at spatial infinity is precisely the Poincar{\'e}
group.  See \cite{AH,A,AR,BS,AshLog,ABR} for further details.

Given the form of the metric (\ref{metric1}), it is natural to
decompose the Einstein equations using the (outward-pointing) unit
vector $n^a = N^{-1}\,\delta^{a}_{\,\,\rho}$ normal to the constant $\rho$ surface $\mathcal{H}_{\rho}$ with induced metric $h_{ab} = g_{ab} -n_a n_b$. Denoting projection onto $\mathcal{H}_{\rho}$ by $ \bot $, this yields
 \begin{eqnarray}
 \bot \left(R _{ab}\right) &=&\mathcal{R}_{ab}+\mathcal{D}%
_{a}a_{b}-a_{a}a_{b}-\pounds
_{n}K_{ab}-KK_{ab}+2K_{a}^{\phantom{a}c}K_{db},
\label{ee1} \\
 \bot\left( R_{a c }n^{c }\right)  &=&\mathcal{D}^{b}K_{ab}-\mathcal{D}_{a}K=-\mathcal{D}^{b}\pi _{ab},
  \label{ee2} \\
R_{a b }n^a n^{b} &=&-\pounds _{n}K-K^{ab}K_{ab}+\left( \mathcal{D}%
_{b}a^{b}-a^{b}a_{b}\right),    \label{ee3}
\end{eqnarray}%
where $\mathcal{R}_{ab}$ is the intrinsic Ricci tensor on $\mathcal{H}_{\rho}$, $\mathcal{D}_{a}$ is the
(torsion-free) covariant derivative compatible with the induced metric, and $\pounds _{v}$ denotes the Lie derivative along a vector field $v^a$.  The
`acceleration' $a^b$ and extrinsic curvature $K_{ab}$ are
 \be a^{b }
= \pounds_{n} n^{b} \qquad K_{ab}=\frac{1}{2}\pounds _{n}h_{ab} ~.
\label{ee4}
 \ee
One may take the fundamental variables to be the metric
$h_{ab}$ on ${\cal H}_{\rho}$ and its conjugate momentum $
\pi^{ab}=h^{ab}K-K^{ab}$.

We now discuss the asymptotic expansions of equations (\ref{ee1}) - (\ref{ee3}),
taking care to distinguish between the $d=4$ and $d>4$ cases.

\subsubsection*{Asymptotic Expansion: \mbox{\boldmath $d=4$}}

The leading term in the asymptotic expansion of the field equation (\ref{ee1}) implies
that the three dimensional metric $h_{ab}^{_{(0)}}$ is a solution of
\begin{equation}
  \mathcal{R}_{ab}^{(0)} = 2 \, h_{ab}^{(0)} ~.
\end{equation}
By inserting the expansion (\ref{metric2}) into the field equations,
Beig \cite{B} showed that the first order Einstein equations are
identically satisfied if
\begin{equation}
D^2 \s + 3 \,\s =0. \label{sigma}
\end{equation}
Here we have introduced the (torsion-free) covariant derivative
$D_a$ on the hyperboloid compatible with the metric $h^{_{(0)}}_{ab}$.
Beig \cite{B} also showed that the second order equations may be written in the form
\begin{eqnarray}
\label{beig20a}h^{(2)}_a{}^a &=& 12 \,\s^2 + \s_a \s^a, \\
\label{beig20b} D_b h^{(2)}_a{}^{b} &=& 16 \,\s \s_a + 2 \,\s_b \s^b_a, \\
\label{beig20c}D^2 h^{(2)}_{ab} - 2 \,h^{(2)}_{ab} &=& 6 \,\s_c \s^c
h^{(0)}_{ab} + 8 \,\s_a \s_b + 14 \,\s \s_{ab} - 18 \,\s^2 h^{(0)}_{ab} +
2 \,\s_{ac} \s^c_b + 2 \,\s_{abc} \s^c,
\end{eqnarray}
where $\s_a = D_a \sigma$, $\s_{ab} = D_b D_a \sigma$, and $\s_{abc}
= D_c D_b D_a \s$ . For further details the reader is referred to \cite{BS, B, MMV}.

\subsection*{Asymptotic Expansion: \mbox{\boldmath $d>4$}}

It is straightforward to carry out calculations analogous to those
of \cite{B} for $d > 4$.  The details are presented in appendix
\ref{section:a0}; a discussion can also be found in \cite{skenAF}. The equations obtained from the asymptotic expansion of (\ref{ee1}) are
\begin{eqnarray}\label{Rab0}
\mathcal{R}_{ab}^{{(0)}} & = & \left( d-2\right) h_{ab}^{{(0)}} \\ \label{Rab1}
\mathcal{R}_{ab}^{{(1)}} & = & D_{a}D_{b} \sigma - \left(d-1\right) \sigma \, h_{ab}^{{(0)}}
    - \frac{d-3}{2} h^{{(1)}}h_{ab}^{{(0)}} + \frac{\left(d-1\right)}{2}\,h_{ab}^{{(1)}} \\ \label{Rab2}
\mathcal{R}_{ab}^{{(2)}} & = & \left( d-2\right) \,\left(  h_{ab}^{{(2)}} - \frac{1}{2}\,h^{{(2)}}\,h_{ab}^{{(0)}} \right)~,
\end{eqnarray}
and the expansion of (\ref{ee2}) yields
\begin{eqnarray}\label{ee11a}
 D^{b} h_{ab}^{(1)} - D_{a} h^{(1)} & = & 2\,\left(\frac{d-2}{d-3}\right)\,D_{a} \sigma \\
 D^{b} h_{ab}^{(2)} - D_{a} h^{(2)} & = & 0 ~. \label{ee11}
\end{eqnarray}
The function $\sigma$ satisfies an equation similar to (\ref{sigma}), which is obtained by taking the trace of (\ref{Rab1}) with $h^{ab}_{^{(0)}}$ and using the covariant divergence of (\ref{ee11a}) to simplify the result. This gives
\begin{equation}
D^{2}\sigma +(d-3)\left(d-1\right) \sigma +\frac{(d-3)\,\left( d-4\right) }{2}%
h^{{(1)}}  = 0 ~.  \label{ee13a}
\end{equation}
In four dimensions the equations for the second order term $h_{ab}^{_{(2)}}$ receive contributions that are quadratic in $h_{ab}^{_{(1)}}$ and $\sigma$, because these terms all appear at the same order in the asymptotic expansions. This is no longer the case in $d>4$, and as a result the analog of (\ref{beig20a}) is
\begin{eqnarray}
h^{{(2)}} &=&0 \: .  \label{ee13b}
\end{eqnarray}%
Note that in the asymptotic expansion of Minkowski space the metric $h^{_{(0)}}_{ab}$ is a metric for the maximally symmetric hyperboloid, i.e., for de Sitter space.  We will only require $h^{_{(0)}}_{ab}$ to be a general Einstein metric satisfying (\ref{Rab0}). However, we will continue to  refer to $\mathcal{H}$ as the `hyperboloid', for convenience.

\subsection{Electric Part of the Weyl Tensor}
The electric part of the Weyl tensor figures prominently in several of our results. It is defined as
\begin{equation}
  E_{ab} := C_{acbd} \, n^{c} n^{d} ~.
\end{equation}
On a constant $\rho$ surface $\mathcal{H}_{\rho}$ this can be
given a purely geometric expression in terms of the intrinsic and
extrinsic curvatures, the acceleration $a^{b}$, and the covariant derivative $\mathcal{D}_{a}$
\begin{eqnarray}
 E_{ab} & = & \left(\frac{d-3}{d-2}\right)\,\left(\frac{1}{d-1}\,h_{ab}\,\pounds_{n} K  - \pounds_{n}K_{ab}\right)
    + \left(\frac{d-4}{d-2}\right)\,K_{a}^{\,\,\, c}\,K_{b c}\\ \nonumber
    & &  + \frac{1}{d-1}\,h_{ab}\,K^{cd}\,K_{cd}     + \frac{1}{d-2}\,K \, \left(K_{ab}- \frac{1}{d-1}\,h_{ab}\,K \right) - \frac{1}{d-2}\,\left(\mathcal{R}_{ab}
    - \frac{1}{d-1}\,h_{ab}\,\mathcal{R} \right) \\ \nonumber
    & & + \left(\frac{d-3}{d-2}\right)\,\left(\mathcal{D}_{a} a_{b} - \frac{1}{d-1}\,h_{ab} \,\mathcal{D}_{c}\, a^{c}
    -a_{a}\,a_{b} + \frac{1}{d-1}\,h_{ab}\,a_{c}\,a^{c}\right) ~.
\end{eqnarray}
It is straightforward to verify that $E_{ab}$ is traceless and
divergenceless. In $d\geq 4$ dimensions the asymptotic expansion takes the form
\begin{equation}\label{EWeylExpansion}
E_{ab} = \rho^{3-d}\,E_{ab}^{(1)} + \rho^{2-d}\,E_{ab}^{(2)} + \mathcal{O}(\rho^{1-d}) ~.
\end{equation}
However, there are differences between the terms in this expansion in the $d=4$ and $d>4$ cases. In four dimensions they are given by
\begin{eqnarray} \label{EWeyl14D}
  E_{ab}^{(1)} & = & - \sigma_{ab} - h_{ab}^{(0)}\,\sigma \\ \label{EWeyl24D}
  E_{ab}^{(2)} & = & -h_{ab}^{(2)} + \s \, \s_{ab} + 5\,h_{ab}^{(0)}\,\s^{2} - 2\,\s_{a}\,\s_{b} + h_{ab}^{(0)}\,\s_{c}\,\s^{c} ~,
\end{eqnarray}
while in $d>4$ dimensions the terms in the expansion are
\begin{eqnarray} \label{EWeyl1}
  E_{ab}^{(1)} & = & - \frac{(d-3)(d-4)}{2}\,h_{ab}^{(1)} - \s_{ab} - (d-3)\,h_{ab}^{(0)}\,\s \\ \label{EWeyl2}
  E_{ab}^{(2)} & = & - \frac{(d-3)(d-2)}{2}\,h_{ab}^{(2)} ~.
\end{eqnarray}
The structure of these terms and the role they play in the construction of conserved charges are discussed in detail in section \ref{sec:CC}.

In four dimensions the second order term (\ref{EWeyl24D}) satisfies a non-trivial identity
 \be \label{ElectricMagnetic}
 E^{_{(2)}}_{ab} - \s E^{_{(1)}}_{ab}= \epsilon_{cd(a}D^c \beta_{b)}^{\,\,d},
 \ee
where $\beta_{ab}$ is the second order magnetic part of the Weyl tensor \cite{B, AH, A, MMV}. This identity is useful in establishing the equivalence between our expressions for the Lorentz charges and the results of Ashtekar and Hanson \cite{AH}.

\setcounter{equation}{0}
\section{A Variational Principle and the Boundary Stress Tensor}
\label{varprinc}

It was shown in \cite{MM} that a good variational principle for gravity
with asymptotically flat boundary conditions in $d \ge 4$ dimensions is
given by the action
\begin{equation}
\label{covaction} S = \frac{1}{16\pi G} \int_{\cal M} \sqrt{-g}\,R +
\frac{1}{8\pi G} \int_{\partial {\cal M}} \sqrt{-h} \,(K - \hat K),
\end{equation}
where $\hat K := h^{ab} \hat K_{ab}$ and $\hat K_{ab}$ is defined to satisfy
\begin{equation}\label{DefEqn}
\label{Khat} {\cal R}_{ab} = \hat K_{ab} \hat K - \hat K_a^{~c} \hat
K_{cb} ~.
\end{equation}
This equation (\ref{Khat}) admits more than one solution for
$\hat{K}_{ab}$ -- we choose the solution that asymptotes to the
extrinsic curvature of the boundary of Minkowski space as $\partial
{\cal M}$ is taken to infinity. As described in \cite{MM}, the
boundary term can be motivated from the heuristic idea that one should subtract off a
``background'' divergence. The defining equation (\ref{Khat}), arrived at via the Gauss-Codazzi
equations, identifies $\hat{K}_{ab}$ as the part of the extrinsic curvature
that is fixed by the kinematics of the theory.  A similar boundary term has been used
to construct variational principles for asymptotically linear dilaton spacetimes
\cite{DonAmitabh, BobRobb} and asymptotically plane wave spacetimes
\cite{Ross}.

Varying the action and imposing the equations of motion yields
\begin{equation}\label{ActionVariation}
\delta S = \frac{1}{16 \pi G} \, \int_{\partial M} \nts\nts \sqrt{-h}\,\left(\pi^{ab} - \hat{\pi}^{ab} + \Delta^{ab} \right)\delta h_{ab}
\end{equation}
where $\hat{\pi}^{ab} \defeq h^{ab}\,\hat{K} - \hat{K}^{ab}$, in analogy with $\pi^{ab}$, and $\Delta^{ab}$ represents a number of `extra' terms. An expression for $\Delta_{ab}$ is given in appendix \ref{sec:ExplicitForm}. The boundary stress tensor is defined as the functional derivative of the on-shell action with respect to $h_{ab}$
\begin{equation}\label{StressTensor}
T^{ab} \defeq \frac{2}{\sqrt{-h}}\,\frac{\delta S}{\delta h_{ab}} = \frac{1}{8 \pi G}\,\left(\pi^{ab} - \hat{\pi}^{ab} + \Delta^{ab} \right)
\end{equation}
The first two terms in (\ref{StressTensor}) were considered in \cite{MM,MMV}
 \be
 T_{ab}^{\Delta\pi}=\frac{1}{8\pi G} \left( \pi_{ab} -
\hat{\pi}_{ab}\right) ~.
 \ee
These terms are conserved in the sense that $\mathcal{D}^{b} T_{ab}^{\Delta \pi} = 0$, and the generators of asymptotic symmetries constructed using $T_{ab}^{\Delta \pi}$ agree with the Ashtekar-Hansen charges. However, the last term in (\ref{StressTensor}) was overlooked in previous investigations of the stress tensor. In the next section we show that this does not alter the conclusions of \cite{MM,MMV}.

\setcounter{equation}{0}
\section{Conserved Charges}
\label{sec:CC}

By the general arguments given in \cite{BY,MM,HIM}, the boundary stress tensor leads to conserved charges that generate the expected asymptotic symmetries.  The generator of diffeomorphisms along an asymptotic Killing field
$\xi^a$ is
 \bea
  Q[\xi] &:=& \lim_{\rho \rightarrow
\infty} \int_{C_\rho} \nts \sqrt{-h_{C_\rho}} \, u_{C_\rho}^a  T_{ab} \, \xi^b
\label{Q1},
 \eea
where each $C_\rho$ is a Cauchy surface within a constant $\rho$ hypersurface ${\mathcal H}_\rho$ such that $C = \lim_{\rho \rightarrow \infty} C_\rho$ is a Cauchy surface
in the boundary hyperboloid ${\mathcal H}$, and
$u^a_{C_\rho}$ is the unit normal to $C_\rho$ in ${\mathcal H}_\rho$.
Explicit computations in \cite{MM, MMV} showed that for $d=4$ the quantity
 \bea
  Q[\xi] &:=& \lim_{\rho \rightarrow
\infty} \int_{C_\rho} \nts \sqrt{-h_{C_\rho}} \, u_{C_\rho}^a  T^{\Delta \pi}_{ab}  \xi^b
 \label{QD},
 \eea
computed without the term $\Delta_{ab}$, agrees precisely with the
Ashtekar-Hansen conserved quantities \cite{AH}. In section \ref{bst4} we show
that the extra terms in $T_{ab}$ do not change this result. The conserved charges do
not receive contributions from $\Delta_{ab}$, despite the fact that in $d=4$ these terms are non-zero
at the relevant order in the asymptotic expansion. The proof relies on a
detailed calculation of $\Delta_{ab}$ in appendix \ref{bstd=4}.
The $d > 4$ case is treated in section \ref{bstd}. A calculation of $\Delta_{ab}$
in appendix \ref{astdg4} shows that the extra terms are identically zero for $d>4$, up to
terms that vanish too fast in the limit $\rho \to \infty$ to
contribute to conserved charges.

\subsubsection*{Asymptotic Killing Fields}

Recall that Minkowski space Killing fields naturally divide into
two classes: translations and Lorentz transformations. In an asymptotically flat spacetime there is a one-to-one correspondence between these classes and the first two terms in the asymptotic expansion of a general asymptotic Killing field $\xi^{a}$. In dimension $d \geq 4$ this expansion takes the form
\begin{equation}\label{KVExpansion}
 \xi^a = \xi_{(0)}^{a} +
 \rho^{-1}\,\xi_{(1)}^{a} + \mathcal{O}(\rho^{-2}) ~.
\end{equation}
Spatial rotations and boosts are associated with the leading order term in the expansion. They correspond to Killing vectors $\xi^{a}_{^{(0)}}$ of $\mathcal{H}$ and therefore satisfy $D_{(a} \xi^{_{(0)}}{\!}_{b)} =0$. Translations appear at next-to-leading order in (\ref{KVExpansion}), and are associated with conformal Killing vectors $\xi_{^{(1)}}^{a}$ that satisfy $D_{(a} \xi^{_{(1)}}{\!}_{b)} =f(\eta) h^{_{(0)}}_{ab}$ for some non-zero function $f$ on $\mathcal{H}$. Terms of order $\rho^{-2}$ and higher are not relevant for our purposes -- the
asymptotic expansion of the stress tensor implies that they do not
contribute to the conserved charges.

In $d=4$ the surface $\mathcal{H}$ at spatial infinity is the unit hyperboloid, so the meaning of asymptotic translations and Lorentz transformations is clear.
However, in dimension $d>4$ we only require $h^{_{(0)}}_{ab}$ to be a general Einstein metric satisfying (\ref{Rab0}). In that case the properties described above, which depend on the isometries and conformal isometries of the metric at spatial infinity, define what we mean by ``translations'' and ``Lorentz transformations.''

\subsubsection*{Potentials for the Stress Tensor}

We now pause to establish a few useful results concerning potentials for transverse, symmetric tensors. The existence of these potentials plays an important role in the construction of the conserved charges. Much of our treatment follows that of \cite{AH,A, B}.

A tensor $\theta_{ab}$ is said to admit a {\it scalar potential} $\alpha$ if
 \be \label{T2ScalarPot}
 \theta_{ab} = D_a D_b \,\alpha - h_{ab}^{(0)} \, D^2 \alpha - (d-2)\,\alpha\, h_{ab}^{(0)} ~,
 \ee
where the metric $h_{ab}^{_{(0)}}$ satisfies (\ref{Rab0}).
Taking the divergence of $\theta_{ab}$ and commuting covariant derivatives shows that it is conserved: $D^a \theta_{ab}=0$. If $\xi_{^{(0)}}^{a}$ is a Killing vector of $h_{ab}^{_{(0)}}$ then the current $\theta_{ab} \, \xi_{^{(0)}}^{b}$ can be expressed as the divergence of an anti-symmetric tensor
\be \label{T2pot}
  \theta_{ab} \,\xi_{(0)}^{b} = D^b\left( 2 \,\xi^{(0)}{\!}_{[b} \,D_{a]}\alpha  + \alpha \,D_{[b} \,\xi^{(0)}{\!}_{a]}\right) ~.
\ee
Now consider a Cauchy surface $C \subset \mathcal{H}$, with timelike unit vector $u^{a}_{(0)}$. The component of the current along $u^{a}_{(0)}$ becomes a total derivative on $C$
\begin{eqnarray}
  u^a_{(0)} \,\theta_{ab} \,\xi_{(0)}^{b}  &=&  u^{a}_{(0)} D^{b}\left( 2 \,\xi^{(0)}{\!}_{[b} \,D_{a]}\alpha
  + \alpha \,D_{[b} \,\xi^{(0)}{\!}_{a]}\right)  \\ \nonumber
  & = &   D_{_{C}}^a\left[\left( 2 \,\xi^{(0)}{\!}_{[a} \,D_{b]}\alpha + \alpha \,D_{[a} \,\xi^{(0)}{\!}_{b]}
  \right)u^b_{(0)} \right] ~.
\end{eqnarray}
This vanishes when integrated over $C$, so that currents of the form $\theta_{ab}\,\xi_{(0)}^{b}$ do not contribute to the conserved charge associated with $\xi_{^{(0)}}^{a}$.

Similarly, a tensor $t_{ab}$ is said to admit a {\it symmetric,  transverse tensor
potential} if
\be \label{TensorPotential}
 t_{ab} = D^2 \gamma_{ab} + 2 \,\mathcal{R}^{(0)}_{acbd}\,\gamma^{cd}
\ee
for some symmetric $\gamma_{ab}$ with $D^{a} \gamma_{ab} = 0$. The tensor $t_{ab}$ is conserved, and for $\xi_{^{(0)}}^{a}$ a Killing field of $h^{_{(0)}}_{ab}$ the current $t_{ab}\,\xi_{^{(0)}}^{b}$ is the divergence of an anti-symmetric tensor
\begin{equation}
  t_{ab} \, \xi_{(0)}^{b} =  2\,D^{a} ( \xi_{(0)}^{c}\,D_{[a} \gamma_{b]c} + \gamma_{c[a}D_{b]} \xi_{(0)}^{c}) ~.
\end{equation}
A current of this form does not contribute to conserved charges, because the component along $u^{a}_{(0)}$ is a total divergence on $C$. The proofs of these statements, which are more involved than in the case of the scalar potential, allow for metrics with non-vanishing Weyl curvature.

\subsection{$d=4$}
\label{bst4}
In this section we show that $\Delta_{ab}$
does not contribute to the conserved charges in four dimensions. That is,
 \be
 \Delta Q[\xi] := \lim_{\rho \rightarrow
\infty} \frac{1}{8\pi G}\int_{C_{\rho}} \nts \sqrt{-h_{C_\rho}} \, u_{C_\rho}^{a}  \Delta_{ab}\,\xi^{b}  = 0 \label{charge}
 \ee
for any asymptotic Killing field $\xi^{a}$. We show that $\Delta Q[\xi]$ is zero using the asymptotic expansion of $\Delta_{ab}$. The main result of this calculation, carried out in detail in appendix \ref{bstd=4}, is equation (\ref{final})
\bea
\Delta_{ab} &=:& \rho \, \Delta^{(0)}_{ab} + \Delta^{(1)}_{ab} + \rho^{-1}\Delta^{(2)}_{ab} + \mathcal{O}(\rho^{-2}) \nn \\
&=&  \frac{1}{4\rho}\,
\left[
\,9 \,\s_c \s^c h^{(0)}_{ab} -29 \,\s_a \s_b + 63 \,\s\s_{ab} + 24 \,\s_{ap}\s_{b}{}^{p} -5 \,\s_{cd}\s^{cd}\,h^{(0)}_{ab} + 45 \,\s^2 \,h^{(0)}_{ab}\right. \nn \\  & & \left. \quad - \,  3\,\s_{mnp}\s^{mnp}\,h^{(0)}_{ab} + 9 \,\s_{pq(a}\s^{pq}{}_{b)} -3 \,\s^{pq}\s_{pq(ab)} - 2\,\s^e\s_{e(ab)}
\right] + \mathcal{O}(\rho^{-2}) ~.
\eea
Three remarks are in order here. First, both $\Delta^{_{(0)}}_{ab}$ and $\Delta^{_{(1)}}_{ab}$ are identically zero. Second, all of the terms in $\Delta_{ab}^{_{(2)}}$ involving $h_{ab}^{_{(2)}}$ cancel, so that it depends only on $\s$ and its derivatives. Third, $\Delta_{ab}^{_{(2)}}$ is traceless: $h^{{(0)} ab} \Delta_{ab}^{_{(2)}} = 0$.

In \cite{MMV} it was  shown that  $ \Delta \pi_{ab} := \pi_{ab} - \hat \pi_{ab}$ is conserved. From general arguments given in \cite{MM,HIM} we expect the full $T_{ab}$ to be conserved as well. It follows that $\Delta_{ab}$ should  be independently conserved. Using various identities from the appendix \ref{collection} one can verify that the divergence of $\Delta_{ab}$ is indeed zero: $\mathcal{D}^a\Delta_{ab}= D^a\Delta^{(2)}_{ab}=0$. Thus, any contributions to the charges from (\ref{charge}) would also be conserved. Since $\Delta^{_{(1)}}_{ab}=0$ there are no contributions to the charges associated with asymptotic translations. There may, however, be a contribution to the Lorentz charges from the non-vanishing term $\Delta_{ab}^{_{(2)}}$
\be \Delta Q[\xi] =   \frac{1}{8 \pi G} \, \int_{C} \nts \sqrt{-h_{C}^{_{(0)}}} \,
 u_{(0)}^{a} \,\Delta^{(2)}_{ab} \, \xi_{(0)}^{b}  ~,
\ee
where the integral is performed over a cut $C$ of the unit hyperboloid $\mathcal{H}$. In the rest of this section we will only be concerned with Lorentz charges constructed from asymptotic
Killing fields with $\xi_{^{(0)}}^{a} \neq 0$.

Most of the terms in $\Delta^{(2)}_{ab}$ can be expressed in terms
of the scalar and tensor potentials described above. An explicit calculation
shows that
\be 4 \,\Delta^{(2)}_{ab} = \theta_{ab} + t_{ab}  + \frac{1}{3}
\,\gamma_{ab} + \frac{11}{2} \,j_{ab} - 4 \,k_{ab} \label{decompose}
\ee
where $\theta_{ab}$ and $t_{ab}$ were defined in (\ref{T2ScalarPot}) and (\ref{TensorPotential}), respectively, and
\bea
\alpha &=& -\frac{3}{2} \,\s_{mn}\s^{mn} - \s^m \s_m, \\
\gamma_{ab} &=& 6 \,\left( 2 \,\s \s_{ab} + \frac{5}{2} \,\s^2 h^{(0)}_{ab} + \s_{a}{}^{c}\s_{cb} - \frac{1}{2} \,\s_{mn}\s^{mn}\,h^{(0)}_{ab}\right), \label{gamma1}\\
 k_{ab} &=& - \frac{1}{2} \,\s^k \s_k \,h^{(0)}_{ab} + \s_a \s_b + \frac{3}{2}\, \s^2\, h^{(0)}_{ab}, \label{kdef} \\
 j_{ab} &=& 2 \,\left(\s \,\s_{ab} + 2 \,\s^2\, h^{(0)}_{ab} + \s_a \s_b - \s_k \s^k \,h^{(0)}_{ab}\right) ~.
\eea
Note that $k_{ab}$, $j_{ab}$, and $\gamma_{ab}$ are all divergence free. Furthermore, $\gamma_{ab}$ and $j_{ab}$ themselves admit similar potentials \cite{MMV}
\bea
\frac{1}{6} \,\gamma_{ab} \,\xi_{(0)}^{a} &=& D^a \left(
\xi_{(0)}^{c} D_{[a} k_{b]c} + k_{c[a} D_{b]} \xi_{(0)}^{c} \right) +  j_{ab}\,\xi_{(0)}^{a}, \label{gamma}\\
 j_{ab}\,\xi_{(0)}^{a}  &=& D^a \left( \s^2 D_{[a} \, \xi^{(0)}{\!}_{b]} - 4\, \s
\s_{[a} \xi^{(0)}{\!}_{b]}\right) ~.
\eea
Thus, with the possible exception of $k_{ab}$, the terms in (\ref{decompose}) cannot contribute to the conserved charges.

We will now show that the $k_{ab}$ term does not contribute to the conserved charges. First, consider the case of spatial rotations. It is convenient to introduce
coordinates $(\tau, \theta, \phi)$ on the unit hyperboloid ${\cal
H}$ so that the metric takes the standard form
 \be
 h^{(0)}_{ab}dx^a dx^b = - d\tau^2 + \cosh^2 \tau \,(d\theta^2 + \sin^2
\theta \,d\phi^2)~.
 \ee
Without loss of generality the cut $C$ is taken to be the surface $\tau = 0$, or ``neck", of
the hyperboloid, which is just the unit two-sphere. In that case the timelike vector $u_{^{(0)}}^{a}$ points in the $\tau$-direction, and for any rotational Killing field $\xi_{^{(0)}}^{a}$ we have
\be
  u_{(0)}^{a} \, h^{(0)}_{ab}\, \xi_{(0)}^{b} = 0 ~.
\ee
Thus, the contribution to the conserved charge from the $k_{ab}$ term in (\ref{decompose}) is given by the integral of $\s_{\tau} \s_{a}\, \xi_{^{(0)}}^{a}$ over the neck of the hyperboloid
\be
Q_{k} = \int_{S^{2}}  \s_\tau \, \s_a \,\xi_{(0)}^{a} ~.
\ee
Since $\s$ has even parity on $\mathcal{H}$ it is convenient to decompose
it as
 \be
 \s = T_{+} Y_{+} + T_{-} Y_{-}
  \ee
where $Y_{\pm}$ schematically denotes even (odd) parity spherical
harmonics on $S^2$, and $T_{\pm}$ is an even (odd) parity function
of the coordinate $\tau$ alone. Note that $T_{-}$ is identically
zero at $\tau=0$. Using this decomposition we have
 \bea
  \s_\tau = \dot T_{+} Y_{+} + \dot T_{-} Y_{-}
 \eea
and
 \bea
 \s_a \,\xi_{(0)}^{a} = T_{+} Y_{+ \xi} + T_{-} Y_{-\xi}~,
  \eea
where $Y_{\pm \xi} = \xi_{(0)}^{a} D_a Y_{\pm}$. Thus,
 \be
 \label{429}
 Q_{k} = (\dot
T_{-}T_{+})  \int_{S^{2}} Y_{+ \xi} Y_{-} ~.
 \ee
But $Y_{+ \xi} $ has even parity when $\xi_{^{(0)}}^{a}$ is a spatial rotation, so the integrand in (\ref{429}) is odd and $Q_k$ vanishes.

A similar argument holds for boosts. On the $\tau=0$ slice of the
hyperboloid, a boost Killing field $\xi_{^{(0)}}^a$ can be expressed
as\,\footnote{On a different cut the vector will have other
components.}
\be
  \xi_{(0)}^{a} = g \, u_{(0)}^{a}
\ee
where $g$ is a function on $S^{\,2}$ with odd parity \cite{AM2}. Thus,
\be
 Q_{k} = \int_{S^{2}} u_{(0)}^{a} k_{ab} \,\xi_{(0)}^{b} = \int_{S^{2}} u^a_{(0)} k_{ab} \, u^b_{(0)}\,g = \int_{S^{2}} k_{\tau \tau} \,g ~. \label{int}
 \ee
From the definition (\ref{kdef}) we see that $k_{\tau \tau}$ is a sum of squares, and
therefore has even parity on $S^{\,2}$. But since $g$ has odd
parity on $S^{\,2}$, the integrand in (\ref{int}) is again odd and
$Q_k$ vanishes. This completes the proof that the extra term
$\Delta_{ab}$ in the stress tensor does not contribute to the
conserved charges in four dimensions.

\subsubsection*{Asymptotic Expansion of the Full Stress Tensor}

We close this section by summarizing our results for the boundary stress tensor in four dimensions. The asymptotic expansion of the full stress tensor is
 \be
T_{ab} = \frac{1}{8 \pi G} \left[\Delta \pi^{(1)}_{ab} + \frac{ \Delta \pi^{(2)}_{ab} + \Delta^{(2)}_{ab}}{\rho} + \ldots \right].
 \ee
The large $\rho$ expansion of $\Delta \pi_{ab}$ was performed in detail in appendix B of \cite{MMV}, where
it was shown that

\bea \label{EWeyl1}
  \Delta \pi^{(1)}_{ab} &= & \s_{ab} + \s \,h^{(0)}_{ab} \,\, = \,\, - E^{(1)}_{ab}  \\ \nn
  \Delta \pi^{(2)}_{ab} &=& - \left( \frac{5}{2} \,\s^2 + \s_c \s^c +
  \frac{1}{2} \,\s_{cd} \s^{cd} \right) h^{(0)}_{ab} + h^{(2)}_{ab} + 2\,\s_a
  \s_b + \s \s_{ab} + \s_{a}{}^c\s_{cb} \\ \label{DeltaPi2}
   & = & - E_{ab}^{(2)} + \frac{1}{6} \, \gamma_{ab} ~.
\eea
This expansion can be used to extract expressions for the generators of asymptotic symmetries from (\ref{Q1}). The term $\Delta_{ab}$ does not contribute, so the relevant terms are
\begin{equation}
Q[\xi] = -\frac{1}{8\,\pi\,G}\,\lim_{\rho  \to \infty} \int_{C} \nts d^{\,2}x \, \sqrt{-h^{(0)}_{C}}\,\left[\rho\,u^{a}_{(0)}\,E_{ab}^{(1)}\,\xi^{b}_{(0)} + u^{a}_{(0)}\,E_{ab}^{(1)}\,\xi^{b}_{(1)} + u^{a}_{(0)}\,\left( E_{ab}^{(2)} - \sigma\,E_{ab}^{(1)} - \frac{1}{6} \, \gamma_{ab} \right)\,\xi^{b}_{(0)} \right] ~.
\end{equation}
Two of the terms in this expression vanish when integrated over $C$. The integral of the first term vanishes because $\xi^{b}_{^{(0)}}$ is a Killing vector on the hyperboloid and $E_{ab}^{_{(1)}}$ takes the form (\ref{T2ScalarPot}), with scalar potential $\sigma$. The integral of the term involving $\gamma_{ab}$ also vanishes, by virtue of (\ref{gamma}). The remaining terms are
\begin{equation}\label{FinalCharge4D}
Q[\xi] = -\frac{1}{8\,\pi\,G}\, \int_{C} \nts d^{\,2}x \, \sqrt{ - h^{(0)}_{C}}\,\left[ u^{a}_{(0)}\,E_{ab}^{(1)}\,\xi^{b}_{(1)} + u^{a}_{(0)}\, \left( E_{ab}^{(2)} - \sigma\,E_{ab}^{(1)}\right) \,\xi^{b}_{(0)} \right] ~,
\end{equation}
where the $\sigma E^{_{(1)}}_{ab}$ term is due to sub-leading terms in the asymptotic expansions of $\sqrt{-h_C}$ and $u^a$.
The charges for asymptotic translations come from the first term in the integrand, and the Lorentz charges come from the second term. The equivalence between the stress tensor and Ashtekar-Hanson definition of the Lorentz charges follows from the identity (\ref{ElectricMagnetic}), applied to the second term in (\ref{FinalCharge4D}) \cite{MMV}.

\subsection{$d>4$}
\label{bstd}

The asymptotic expansion of $\Delta_{ab}$ in dimension $d>4$ takes the form
 \be
 \Delta_{ab} = \rho \,\Delta^{(0)}_{ab} + \rho^{-(d-4)}\,\Delta^{(1)}_{ab} + \rho^{-(d-3)} \, \Delta^{(2)}_{ab}
 + \mathcal{O}(\rho^{-(d-2)}) ~.
 \ee
In appendix \ref{astdg4} we show that $\Delta_{ab}^{_{(0)}} = \Delta_{ab}^{_{(1)}} = \Delta_{ab}^{_{(2)}} =0$, so the stress tensor takes the form
 \bea
T_{ab}&=&\frac{1}{8\pi G} \,\left( \pi_{ab} - \hat{\pi}_{ab}\right) \nonumber \\
&=&  \frac{1}{8\pi G}\,\rho^{\,4-d}\,\left(\sigma \,h^{(0)}_{ab}+\frac{d-4}{2}\,h^{(1)}_{ab} + \frac{1}{d-3}\,
D_{a}D_{b}\sigma \right)  + \frac{1}{8\pi G}\,\rho^{\,3 - d}\,\frac{d-2}{2}\, h^{(2)}_{ab} + \mathcal{O}(\rho^{\,2-d})~. \label{stressdg4}
 \eea
This is a pleasing result because it is proportional to $E_{ab}$, without the additional terms that appeared in four dimensions. The relation between the stress tensor and the electric part of the
Weyl tensor is
\begin{equation}
 T_{ab} = - \frac{1}{8\,\pi\,G}\,\frac{\rho}{d-3}\,E_{ab} + \mathcal{O}(\rho^{2-d}) ~.
\end{equation}
The expression (\ref{Q1}) for the conserved charges becomes
\begin{equation}\label{PreCharge}
Q[\xi] = -\frac{1}{8\,\pi\,G}\,\frac{1}{(d-3)}\,\lim_{\rho  \to \infty} \int_{C} \nts d^{\,d-2}x \, \sqrt{-h^{(0)}_{C}}\,\left[\rho\,u^{a}_{(0)}\,E_{ab}^{(1)}\,\xi^{b}_{(0)} + u^{a}_{(0)}\,E_{ab}^{(1)}\,\xi^{b}_{(1)} + u^{a}_{(0)}\,E_{ab}^{(2)}\,\xi^{b}_{(0)} \right] ~.
\end{equation}
The first term is proportional to $\rho$, and must vanish for the $\rho \to \infty$ limit to exist. In $d=4$ the analogous term vanishes because $E_{ab}^{_{(1)}}$ admits a simple scalar potential. The proof is more complicated for $d>4$, but proceeds in essentially the same way. The term $E_{ab}^{_{(1)}}$ can be expressed in terms of one tensor and two scalar potentials as
\begin{equation}\label{Eab1Decomposition}
  E_{ab}^{(1)} = - \theta_{ab}[\sigma] - \frac{d-4}{2}\,\left( t_{ab} - \theta_{ab}[h^{_{(1)}}] - 2\,\left(\frac{d-1}{d-3}\right)\,\theta_{ab}[\sigma] \right)~,
\end{equation}
with the potential for $t_{ab}$ given by
\begin{equation}
  \gamma_{ab} = h_{ab}^{(1)} - h_{ab}^{(0)} \, h^{(1)} - 2\, \left(\frac{d-2}{d-3} \right)\,h_{ab}^{(0)}\,\sigma ~.
\end{equation}
It follows that the flux across $C$ of the current $E_{ab}^{_{(1)}}\,\xi_{^{(0)}}^{b}$ vanishes. Taking the $\rho \to \infty$ limit gives the final expression for the conserved charges in dimension greater than four
\begin{equation}\label{FinalCharge}
Q[\xi] = -\frac{1}{8\,\pi\,G}\,\frac{1}{(d-3)}\,\int_{C} \nts d^{\,d-2}x \, \sqrt{-h^{(0)}_{C}}\,\left[ u^{a}_{(0)}\,E_{ab}^{(1)}\,\xi^{b}_{(1)} + u^{a}_{(0)}\,E_{ab}^{(2)}\,\xi^{b}_{(0)} \right] ~.
\end{equation}
We remind the reader that the above expressions hold when $h^{_{(0)}}_{ab}$ is any Einstein metric satisfying (\ref{Rab0}). As in the case $d=4$, the charges for asymptotic translations come from the first term and the Lorentz charges come from the second term. However, for $d > 4$ the sub-leading terms in the expansions of $\sqrt{-h_{C_\rho}}$ and $u^a_{C_\rho}$ do not contribute to (\ref{FinalCharge}) as they begin at too high of an order in $\rho^{-1}$.

\setcounter{equation}{0}
\section{Examples}
\label{Examples}
It is useful to illustrate the results of the previous sections with a few concrete examples. In the first example we determine the stress tensor and conserved charges for the Kerr solution in four dimensions. Then, in the second example, we explicitly construct four dimensional spacetimes with $\Delta_{ab}^{_{(2)}} \neq 0$.

\subsection{The Kerr Black Hole}
\label{sec:KerrBH}

The line element for the Kerr solution in four dimensions can be written as
\begin{equation}\label{Kerr1}
ds^{\,2} = \frac{\Xi}{\Delta}\,dr^{2} + \Xi\,d\theta^{\,2} + (r^{2} + a^{2})\,\sin^{2}\theta \,d\phi^{\,2} - dt^{\,2} + \frac{2\,M\,G\,r}{\Xi}\,\left(a\,\sin^{2}\theta \, d\phi - dt \right)^{2} ~.
\end{equation}
The parameters $a$ and $M$ are related to the mass and angular momentum, and the functions $\Xi$ and $\Delta$ are given by
\begin{equation}
  \Xi = r^{2} + a^{2}\,\cos^{2}(\theta) \quad \quad \quad \Delta = r^{2}-2\,M\,G\,r + a^{2} ~.
\end{equation}
The line element must be expressed in the Beig-Schmidt form (\ref{metric1}) before the results of the previous sections can be applied. This is accomplished by a series of coordinate redefinitions, as described in \cite{BS}. In the Beig-Schmidt coordinates, $h_{ab}^{_{(0)}}$ is the standard metric on the hyperboloid
\begin{equation}\label{HyperboloidMetric}
  h_{ab}^{(0)}dx^{a}dx^{b} = -d\tau^{2} + \cosh^{2}\tau \, \left(d\theta^{\,2} + \sin^{2}\theta \, d\phi^{\,2} \right)
\end{equation}
and the function $\sigma$ is given by
\begin{equation}\label{sigmaKerr}
\sigma = M \, G\, \cosh 2\tau \,\, \textrm{sech}\,\tau ~.
\end{equation}
Notice that this is the only admissible solution of (\ref{sigma}) that depends only on $\tau$~\footnote{A second solution is not even under $\tau \to -\tau$, as required by the boundary conditions.}.
The first-order terms in the expansion of the metric are $h_{ab}^{_{(1)}} = -2\,\sigma\,h_{ab}^{_{(0)}}$, as required by the asymptotically flat boundary conditions. The second order terms $h_{ab}^{_{(2)}}$ are a bit more complicated. Their particular form is not too illuminating; it is sufficient to note that $h_{ab}^{_{(2)}}$ is the first term in the asymptotic expansion that depends on the rotation parameter $a$.

The Kerr solution has two non-zero conserved charges: the mass associated with translations $\partial_{t}$ and the angular momentum associated with rotations $\partial_{\phi}$. Calculating the charges from (\ref{FinalCharge4D}) requires the $\tau$-$\tau$ and $\tau$-$\phi$ components of the electric part of the Weyl tensor. Specifically, the terms we need are
\begin{equation}
 E_{\tau\tau}^{(1)} = -2\,M\,G\,\textrm{sech}^{3} \tau \quad \quad \quad E_{\tau\phi}^{(1)} = 0 \quad \quad \quad  E_{\tau\phi}^{(2)} = 3\,a\,M\,G\,\sin^{2}\theta \,\, \textrm{sech}^{2}\tau ~ .
\end{equation}
The mass is obtained by evaluating the first term in (\ref{FinalCharge4D}) for the vector field
\begin{equation}
 \xi^{a} = \rho^{-1} \cosh(\tau) \,\delta^{a}_{~ \tau} + \ldots ~.
\end{equation}
The cut $C$ is taken to be the $\tau=0$ surface of the unit hyperboloid, with timelike normal $u^{a}_{^{(0)}} = \delta^{\mu}_{~\tau}$. Then the conserved charge is
\begin{eqnarray} \nn
 Q_{\partial_{\tau}} & = & - \frac{1}{8 \pi G}\,\int_{C} \nts d^{\,2}x \, \sqrt{-h_{C}^{(0)}} \, \sin \theta \, u^{\tau}_{(0)} \, E_{\tau\tau}^{(1)}\,\xi^{\tau}_{(1)} \\
  & = & \frac{1}{8 \pi G} \, \int_{0}^{2\pi} \nts \nts d\phi \, \int_{0}^{\pi} \nts d\theta \,\cosh^{2}\tau \, \sin \theta \, \cdot 2\,M\,G\,\textrm{sech}^{\,3}\tau \cdot \, \cosh \tau \\ \nn
  & = & M~.
\end{eqnarray}
The vector field for rotations is just $\xi^{a} = \delta^{a}_{~\phi}$. Using this in the second term of (\ref{FinalCharge4D}) gives
\begin{eqnarray} \nn
Q_{\partial_{\phi}} & = & -\frac{1}{8 \pi G} \, \int_{C} d^{2}x \,\sqrt{-h^{_{(0)}}_{C}}\,u^{\tau}_{(0)}\, \left( E_{\tau\phi}^{_{(2)}} - \s E_{\tau \phi}^{_{(1)}}\right) \,\xi^{\phi}_{(0)} \\
  & = & - \frac{1}{8 \pi G} \,\int_{0}^{2\pi} \nts \nts d\phi\,\int_{0}^{\pi} \nts d\theta \,\cosh^{2}\tau \, \sin\theta\,\cdot 3\,a\,M\,G\,{\rm sech}^{2}\tau \, \sin^{2}\theta  \\ \nn
  & = & - a \, M ~.
\end{eqnarray}
These are the expected results for the mass and angular momentum of the Kerr solution.

It is interesting to calculate the extra terms $\Delta_{ab}^{_{(2)}}$ for the Kerr solution. Using the expressions for $h_{ab}^{_{(0)}}$ and $\sigma$, all of the components are found to vanish
\begin{equation}\label{VanishingExtraTerms}
 \Delta_{ab}^{(2)} = 0 ~.
\end{equation}
This result also applies for the Schwarzschild and Reissner-Nordstrom solutions. In these cases the results of section \ref{bst4} are superfluous, since the boundary stress tensor reduces to the form considered in \cite{MMV}.

\subsection{Spacetimes with $\Delta_{ab}^{_{(2)}} \neq 0$}
\label{sec:Exotics}

In the previous section we saw that the extra terms vanish for a number of familiar four dimensional spacetimes. However, it is easy to demonstrate the existence of asymptotically flat spacetimes for which the extra terms do not vanish. The condition (\ref{sigma}) is a linear equation, so new solutions may be obtained by superposition. The result (\ref{decompose}), on the other hand, is non-linear, which means that these new solutions will generally have non-vanishing $\Delta_{ab}^{_{(2)}}$.

An explicit example is a spacetime consisting of two black holes related by a boost. For simplicity we will work with the Schwarzschild solution. Since this is recovered from the $a \to 0$ limit of (\ref{Kerr1}) it follows that $\sigma$ for the Schwarzschild metric is also given by (\ref{sigmaKerr}). First, consider a single black hole that is boosted along the z-axis by an amount $\beta$. The function $\sigma$ becomes
\begin{equation}\label{boostedsigma}
 \sigma_{b} = \frac{~1+2\,q^{\,2}}{\sqrt{1+q^{\,2}}}\,M\,G
\end{equation}
with $q$ given by
\begin{equation}
  q = \cosh \beta \, \sinh \tau + \sinh \beta \, \cosh \tau \, \cos \theta ~.
\end{equation}
The boost simply maps one solution of (\ref{sigma}) to another, and does not change the fact that the extra terms vanish. Now consider a spacetime containing two black holes related by the boost described above. The full metric for this spacetime may be complicated. But the first order term in the asymptotic expansion satisfies a linear equation, so it is given by the sum of (\ref{sigmaKerr}) and (\ref{boostedsigma})
\begin{equation}\label{fullsigma}
 \tilde{\sigma} = \sigma + \sigma_{b} ~.
\end{equation}
As expected, the extra terms do not vanish for this solution. Assuming that the boost parameter is small, the first non-zero terms appear at second order in $\beta$
\begin{eqnarray}
\Delta_{\tau\tau}^{(2)} & = & \beta^{\,2} \,F_{1}(\tau,\theta) \\
\Delta_{\tau\theta}^{(2)} & = & \beta^{\,2} \, F_{2}(\tau,\theta)\,\sin 2\theta \, \tanh \tau \\
\Delta_{\theta\theta}^{(2)} & = & \beta^{\,2}\,F_{3}(\tau, \theta) \\
\Delta_{\phi\phi}^{(2)} & = & \beta^{\,2}\,F_{4}(\tau, \theta) \\
\Delta_{\tau\phi}^{(2)} & = & \Delta_{\theta\phi}^{(2)} ~\, = ~\, 0
\end{eqnarray}
where the $F_{n}(\tau,\theta)$ are polynomials in $\cos 2\theta$ and $\cosh 2 \tau$. The result is traceless and divergenceless, and the $\theta$ and $\tau$ dependence of the first two terms is such that they do not contribute to the Lorentz charges.

The functions (\ref{boostedsigma}) and (\ref{fullsigma}) contain all the information needed to determine the energy and momentum of the solution described above. The energy and momentum for the Schwarzschild solution are the same as for Kerr, and for the boosted black hole we have
\begin{equation}
E = M \, \cosh \beta \quad \quad p_{z} = M\,\sinh \beta \quad \quad p_{x} = p_{y} = 0 ~.
\end{equation}
These charges are obtained from the first term in (\ref{FinalCharge4D}), which is linear in $\sigma$. Therefore the energy and momentum for solutions like (\ref{fullsigma}) are given by the sum of the energy and momentum of the constituent solutions.

\setcounter{equation}{0}
\section{Conclusion}
\label{disc}
 The boundary stress tensor (\ref{StressTensor})
obtained from the action (\ref{covaction}) can be used to define a
conserved charge for any asymptotic Killing field $\xi^{a}$
\begin{equation}\label{GenericCharge}
 Q[\xi] = \frac{1}{8 \pi G} \int_{C} \nts d^{\,d-2}x \, \sqrt{- h_{C}} \, u^{a} \, T_{ab} \,
    \xi^{b} ~,
\end{equation}
where $(C,h_C)$ is a Cauchy surface in $(\mathcal{H},h)$, and
$u^{a}$ is a timelike unit vector in $\mathcal{H}$ normal to $C$.
Such charges necessarily generate the asymptotic symmetries of the
theory.

In \cite{MMV} it was shown a slightly different expression, in which
$T^{ab}$ in (\ref{GenericCharge}) is replaced by
\begin{equation}\label{IdealTab}
  T^{\Delta \pi}_{ab} = \frac{1}{8 \pi G}\,\left(\pi_{ab} - \hat{\pi}_{ab}
  \right),
\end{equation}
agrees with other definitions of conserved charges in the
literature.   However, the difference between $\Delta_{ab} = T_{ab}-
T^{\Delta \pi}_{ab}$ consists of a number of `extra' terms whose
asymptotic expansion takes the form
\begin{equation}\label{DeltaExpansion}
\Delta_{ab} = \rho^{4-d}\,\Delta_{ab}^{(1)} +
\rho^{3-d}\,\Delta_{ab}^{(2)} + \mathcal{O}(\rho^{2-d}) ~.
\end{equation}
The first two terms in this expansion appear at just the right
orders to contribute to the conserved charges associated with
asymptotic translations and Lorentz transformations, respectively.
This observation creates a certain tension with the results of
\cite{MM, MMV}, which did not take these extra terms into account.

Our main result was to resolve this issue by proving that the extra
terms do not contribute to the conserved charges. In dimension $d>4$
the first two terms in (\ref{DeltaExpansion}) simply vanish, as a
result of a remarkable series of cancelations. The proof in four
dimensions is more subtle. In that case the first order term
vanishes, but the second order term may be non-zero. However, in
section \ref{bst4} we showed that $\Delta_{ab}^{_{(2)}}$ takes the
form (\ref{decompose}), which implies that it cannot contribute to
the conserved charges. The remaining terms in $T_{ab}$ are of the
form (\ref{IdealTab}), so that the analysis of \cite{MM, MMV}
applies without modification. We find that (\ref{GenericCharge}) can
be expressed in terms of the electric part of the Weyl tensor. In
four dimensions the charges are given by
\begin{equation}\label{ConclusionCharge1}
Q[\xi] = -\frac{1}{8\,\pi\,G}\, \int_{C} \nts d^{\,2}x \, \sqrt{ - h^{(0)}_{C}}\,\left[ u^{a}_{(0)}\,E_{ab}^{(1)}\,\xi^{b}_{(1)} + u^{a}_{(0)}\, \left( E_{ab}^{(2)} - \sigma\,E_{ab}^{(1)}\right) \,\xi^{b}_{(0)} \right] ~,
\end{equation}
while in dimension greater than four they are
\begin{equation}\label{ConclusionCharge2}
Q[\xi] = -\frac{1}{8\,\pi\,G}\,\frac{1}{(d-3)}\,\int_{C} \nts d^{\,d-2}x \, \sqrt{-h^{(0)}_{C}}\,\left[ u^{a}_{(0)}\,E_{ab}^{(1)}\,\xi^{b}_{(1)} + u^{a}_{(0)}\,E_{ab}^{(2)}\,\xi^{b}_{(0)} \right] ~.
\end{equation}
The explicit comparisons made in \cite{MMV} show that these charges are equivalent to both the ADM construction \cite{ADM1} and (in $d=4$) the covariant charges defined by Ashtekar and Hanson \cite{AH}.

\subsubsection*{Acknowledgments}
The authors thank Simon Ross and Julian Le Witt for bringing the
additional terms $\Delta_{ab}$ to their attention and Abhay Ashtekar
for a number of useful discussions related to asymptotic flatness.
AV would also like to thank Aaron Amsel and Ian Morrison for useful
discussions.  RM and RMc are supported by the Perimeter Institute for Theoretical
Physics (PI). Research at PI is supported in part by funds
from NSERC of Canada and MEDT of Ontario. DM and AV are supported in part by the National
Science Foundation under Grants No. PHY05-55669 and by funds from the
University of California. AV would like to thank the Indian
Institute of Technology Kanpur, Harish Chandra Research Institute
Allahabad, Tata Institute of Fundamental Research Mumbai, and Inter
University Center for Astronomy and Astrophysics Pune for
hospitality while this work was being completed.

~\\
\appendix

\noindent{\bf \Large Appendices}\\

Throughout the body of the paper indices $a$,$b$,$c$ from the beginning of the alphabet are used for all tensors. This applies to $d$ dimensional spacetime indices, $d-1$ dimensional indices on $\mathcal{H}$, and $d-2$ dimensional indices on a Cauchy surface $C \subset \mathcal{H}$. In the following sections tensor indices range into the middle and upper parts of the alphabet. This is due to the large number of indexed quantities appearing in some of the calculations. All indices should be interpreted in the same manner as the indices used in the main text.

\setcounter{equation}{0}
\section{Explicit Form of $\Delta_{ab}$}
\label{sec:ExplicitForm}

In this appendix we derive an expression for the extra terms $\Delta_{ab}$ that appear in the stress tensor. Varying the action (\ref{covaction}) and imposing the equations of motion yields%
\begin{equation}
\delta S=\frac{1}{16\pi G}\int_{\partial M}\sqrt{-h} \left[ \left( \pi ^{ab} - h^{ab}
\hat{K} + 2\hat{K}^{ab}\right) \delta h_{ab}-2h^{ab}\delta \hat{K}_{ab} \right] ~.
\label{e4}
\end{equation}%
To simplify this we must compute $\delta \hat{K}_{ab}$.  From the defining equation for $\hat K$ (\ref{Khat}) we have
\begin{equation}
L_{ab}{}^{cd}\delta \hat{K}_{cd}=\delta \mathcal{R}_{ab}+\left(
\hat{K}_{ab}\hat{K}^{cd}-\hat{K}_{a}{}^{c}\hat{K}_{b}{}^{
d}\right) \delta h_{cd}  \label{e5}
\end{equation}
where following \cite{MM} we define
\begin{equation}
L_{ab}{}^{cd}=h^{cd}\hat{K}_{ab}+\delta _{a}^{c}\delta _{b}^{d}\hat{K}
-\delta _{a}^{c}\hat{K}_{b}^{d}-\delta _{b}^{c}\hat{K}_{a}^{d}.  \label{e6}
\end{equation}
Inserting equation (\ref{e5}) into (\ref{e4}) we obtain
\begin{equation}
\delta S=\frac{1}{16\pi G}\int_{\partial M}\sqrt{-h}\left[ \left( \pi ^{ab} - \hat{\pi}%
^{ab}+\hat{K}^{ab}-2\tilde{L}^{cd}\left( \hat{K}_{cd}\hat{K}^{ab}-\hat{K}%
_{c}{}^{a}\hat{K}_{d}{}^{b}\right) \right) \delta h_{ab}-2%
\tilde{L}^{ab}\delta \mathcal{R}_{ab} \right] \label{e7}
\end{equation}%
where we define $\hat{\pi}^{ab}=h^{ab}\hat{K}-\hat{K}^{ab}$
and where
\begin{equation}
\tilde{L}^{ab}\equiv h^{mn}\left( L^{-1}\right) _{mn}{}^{ab}  \label{e8},
\end{equation}%
with $\left( L^{-1}\right) _{mn}{}^{ab}$ defined to be the inverse of the
tensor on the left hand side of (\ref{e5}), i.e.,
\begin{equation}
\delta \hat{K}_{ab}=\left( L^{-1}\right) _{ab}{}^{ij}\left[ \delta \mathcal{R%
}_{ij}-\left( \hat{K}_{ij}\hat{K}^{cd}-\hat{K}_{i}{}^{c}\hat{K}_{j}{}^{d}\right) \delta h_{cd}\right] .  \label{e9}
\end{equation}
We can use the fact that
\begin{equation}
\delta \mathcal{R}_{ij}=-\frac{1}{2}h^{kl}\mathcal{D}_{i}\mathcal{D}%
_{j}\delta h_{kl}-\frac{1}{2}h^{kl}\mathcal{D}_{k}\mathcal{D}_{l}\delta
h_{ij}+h^{kl}\mathcal{D}_{k}\mathcal{D}_{(i}\delta h_{j)l},  \label{e10}
\end{equation}%
where $\mathcal{D}_i$ is the torsion-free covariant derivative compatible with $h_{ab}$. We now perform integrations by parts to write
\begin{eqnarray}
\delta S &=&\frac{1}{16\pi G}\int_{\partial M}\sqrt{-h}\Bigg{[}  \pi ^{ab} - \hat{%
\pi}^{ab}+\hat{K}^{ab}-2\tilde{L}^{cd}\left( \hat{K}_{cd}\hat{K}^{ab}-\hat{K}%
_{c}{}^{a}\hat{K}_{d}{}^{b}\right)   \nonumber \\
&& + \mathcal{D}^{2}\tilde{L}^{ab}+h^{ab}\mathcal{D}_{k}\mathcal{D}_{l}%
\tilde{L}^{kl}-\mathcal{D}_{k}\left( \mathcal{D}^{a}\tilde{L}^{kb}+\mathcal{D}^{b}\tilde{L}^{ka}\right) \Bigg{]} \delta h_{ab}.  \label{e11}
\end{eqnarray}%
The boundary stress-tensor for pure gravity can be readily obtained
from this equation
 \be
  T^{ab} = \frac{2}{\sqrt{-h}} \frac{\delta S}{\delta h_{ab}} = \frac{1}{8\pi G} \left( \pi^{ab} -
\hat{ \pi}^{ab} + \Delta^{ab} \right)
 \label{full}
 \ee
where $\Delta_{ab}$ is defined to
be
 \be
  \label{main}
\Delta_{ab}= \hat{K}_{ab}-2\tilde{L}^{cd}(\hat{K}_{cd}\hat{K}_{ab}-\hat{K}%
_{ca}\hat{K}_{bd}) + \mathcal{D}^{2}\tilde{L}_{ab}+h_{ab}\mathcal{D}_{k}\mathcal{D}_{l}%
\tilde{L}^{kl}-2 \mathcal{D}_{k}
\mathcal{D}_{(a}\tilde{L}^{k}{}_{b)}.
 \ee
Some explicit results were given in \cite{AstMannStelea}. The result (\ref{main}) is analyzed in detail for the $d=4$ and $d>4$ cases in sections \ref{bstd=4} and \ref{astdg4}, respectively.

\setcounter{equation}{0}
\section{Stress Tensor Calculation $d=4$}
\label{bstd=4}
This appendix contains  details of the boundary stress calculations in $d=4$.  In particular, we calculate $\Delta_{ab}$ as an expansion in inverse powers of $\rho$. We work with zero NUT charge (i.e., we use $h^{(1)}_{ab} = -2 \sigma h^{(0)}_{ab}$) and make use of asymptotic field equations (\ref{sigma}-\ref{beig20c}) as needed. The calculation is organized as follows: in section \ref{section:b1} we calculate the inverse of $L$. We then perform $\rho^{-1}$ expansions of $\tilde L^{ab}$ and $\tilde L^{ab} M_{abcd}$ in section \ref{section:b2}. In section \ref{section:b3} we calculate $\rho^{-1}$ expansions of the derivative terms needed in (\ref{main}). In section \ref{section:b4} we collect various results and present a final expression for $\Delta_{ab}$. Equation (\ref{final}) is the main result of this appendix. Throughout we make use of identities from appendix \ref{collection}.

\subsection{The Inverse of $L_{ab}{}^{cd}$}
\label{section:b1}
Let us begin by calculating the inverse of $L_{ab}{}^{cd}$. The form of the expansions for $L$ and its inverse are
\begin{eqnarray}
L & = & L^{(0)} + \frac{1}{\rho}\,L^{(1)} + \frac{1}{\rho^{\,2}}\,L^{(2)} + \ldots \\
(L^{-1}) & = & (L^{-1})^{(0)} + \frac{1}{\rho}\,(L^{-1})^{(1)} + \frac{1}{\rho^{\,2}}\,(L^{-1})^{(2)}+\ldots
\end{eqnarray}
with the coefficients in the expansion of $L^{-1}$ given by
\begin{eqnarray} \label{L1inv4d}
\left(L^{-1}\right)^{(1)}{\!}_{ij}{}^{pq} & = & - \left(L^{-1}\right)^{(0)}{\!}_{ij}{}^{kl}\, L^{(1)}{}_{kl}{}^{mn}
\left(L^{-1}\right)^{(0)}{\!}_{mn}{}^{pq} \\ \label{L2inv4d}
\left(L^{-1}\right)^{(2)}{\!}_{ij}{}^{pq} & = & -
\left(L^{-1}\right)^{(0)}{\!}_{ij}{}^{kl}\, L^{(2)}{}_{kl}{}^{mn}
\left(L^{-1}\right)^{(0)}{\!}_{mn}{}^{pq} - \left(L^{-1}\right)^{(1)}{\!}_{ij}{}^{kl} \,L^{(1)}{}_{kl}{}^{mn}
\left(L^{-1}\right)^{(0)}{\!}_{mn}{}^{pq}
\label{L2d4} ~.
\end{eqnarray}
The part of $L^{-1}$ relevant for our calculations is actually $h^{ij}(L^{-1})_{ij}{}^{kl}$. As a first
step we calculate  $h^{{(0)} ij} \left(L^{-1}\right)^{(2)}{}_{ij}{}^{pq}$. Using the expression for
$\left(L^{-1}\right)^{_{(0)}}$ (equation (B.7) of \cite{MM}) and
equation (\ref{L1inv4d}) we obtain
 \bea
 h^{(0)\: ij}
\left(L^{-1}\right)^{(2)}{}_{ij}{}^{pq} &=& - \frac{\rho}{4}\, h^{(0)
\: kl}L^{(2)}_{kl}{}^{mn} \left(L^{-1}\right)^{(0)}{}_{mn}{}^{pq}
\nn \\ & & + \frac{\rho}{4} \,h^{(0) \: ab}L^{(1)}{}_{ab}{}^{cd}
\left(L^{-1}\right)^{(0)}{}_{cd}{}^{kl}L^{(1)}{}_{kl}{}^{mn}
\left(L^{-1}\right)^{(0)}{}_{mn}{}^{pq}~.
 \label{l2}
  \eea
The first term in this equation is linear in $L^{(2)}$, while the second term  is quadratic in $L^{(1)}$.
After a bit of calculation we find
 \bea
 L^{(2)}_{ij}{}^{kl} &=& \frac{1}{\rho} \,\Bigg{[} \, \hat{p}^{(2)}_{ij}
 h^{(0)}{}^{kl} - h^{(1)}{}^{kl} \hat p^{(1)}_{ij} -
(h^{(2)}{}^{kl} - h^{(1)}{}^{kp} h^{(1)}_{p}{}^{l}) h^{(0)}_{ij} \nn
\\ && \quad \,\,\, + \, \delta_{(i}^{k} \delta_{j)}^{l} ( \hat p^{(2)} - \hat
p^{(1)} \cdot h^{(1)}   - h^{(2)}+ h^{(1)} \cdot h^{(1)})  \\ &&
\quad \,\,\,-\, \delta_{i}^{(k} (\hat p^{(2)} {}^{l)}{}_{j} - h^{(1)}{}^{l)m}
\hat p^{(1)}_{jm} - h^{(2)}{}^{l)}{}_{j} + h^{(1)}{}^{l)p}
h^{(1)}_{pj}) \nn \\ & & \quad \,\,\, - \, \delta_{j}^{(k} (\hat p^{(2)}
{}^{l)}{}_{i} - h^{(1)}{}^{l)m}  \hat p^{(1)}_{im} -
h^{(2)}{}^{l)}{}_{i} + h^{(1)}{}^{l)p} h^{(1)}_{pi}) \Bigg{]}~, \nn
 \eea
where $u^{(1)} \cdot v^{(1)} = u^{(1) ij} v^{(1)}_{ij}$. Contracting
with $h^{(0)\: ij}$ yields the first term in (\ref{l2})
 \bea
 h^{(0)\: ij} L^{(2)}_{ij}{}^{kl} &=&
\frac{1}{\rho}\, \Bigg{[} \,2 \,h^{(0)\: kl} \,\hat p^{(2)} - 2 \,\hat p^{(2)
\: kl} - h^{(1)\: kl} \,\hat p^{(1)} - h^{(2) \: kl} + h^{(1)\: kp}
h^{(1)}{}_{p}{}^{l} \nn \\ & & \quad \,\,\, - \,h^{(0)\: kl} \,h^{(1)}\cdot \hat
p^{(1)} - h^{(0)\: kl} \,h^{(2)} + h^{(0)\: kl} h^{(1)} \cdot h^{(1)}
+ 2 \,h^{(1)\: (l|m|} \hat p^{(1) \: k)}{}_{m}\Bigg{]}~.
 \label{l2s}
 \eea
Using $h^{(1)}_{ij} = - 2 \s h^{(0)}_{ij}$ and $\hat p^{(1)}_{ij} =
\s_{ij} - \s h^{(0)}_{ij}$ \cite{MMV} the expression (\ref{l2s})
further simplifies to
 \bea
 h^{(0) \: ij}L^{(2)}_{ij}{}^{mn}
\left(L^{-1}\right)^{(0)}{}_{mn}{}^{pq} &=& h^{(0)\: pq} \,\hat{p}^{(2)} - 2\, \hat{p}^{(2) \: pq} - h^{(2)\: pq} - 4 \,\s^2 h^{(0)\: pq} - 4 \,\s \s^{pq}~.
\label{l2ss}
 \eea
Now consider the second term in equation (\ref{l2}). A short
calculation gives
 \bea
& & \frac{\rho}{4} \,h^{(0) \: ab}L^{(1)}{}_{ab}{}^{cd}
 \left(L^{-1}\right)^{(0)}{}_{cd}{}^{kl}L^{(1)}{}_{kl}{}^{mn}\left(L^{-1}\right)^{(0)}{}_{mn}{}^{pq} \nn \\ & & \quad \quad =  \rho \left[ -\frac{3}{8} \,\sigma_{ab}\s^{ab} h^{(0)\: pq} + \frac{17}{8}\, \s^2 h^{(0)\: pq} + 2 \,\s \s^{pq} + \s^{pl}\s_{l}{}^{q}\right]~.
 \label{l1sss}
 \eea
Adding the two terms (\ref{l2ss} and \ref{l1sss}) one arrives at the
result
 \bea
h^{(0)\: ij} \left(L^{-1}\right)^{(2)}{}_{ij}{}^{pq} & = & \frac{\rho}{4} \Bigg{[} - h^{(0)\: pq} \,\hat p^{(2)} + 2 \,\hat p^{(2) \: pq} + h^{(2)\: pq} + \frac{25}{2}\, \s^2 h^{(0)\: pq} + 12 \,\s \s^{pq} \nn \\ & &  \quad \,\,\,- \,\frac{3}{2} \,\s_{ab}\s^{ab} h^{(0) \: pq} + 4 \,\s^{pl}\s_{l}{}^{q}\Bigg{]}~.\nn \\
\label{linv2}
\eea

\subsection{Expansions of $\tilde L^{ab}$ and $\tilde L^{ab}M_{abcd}$}
\label{section:b2}

In this section we calculate expansions of $\tilde L^{ab}$ and $\tilde L^{ab}M_{abcd}$. Recall that $\tilde L^{ab}$ and $M_{abcd}$ are defined as\bea
\tilde L^{ab} &=& h^{cd}(L^{-1})_{cd}{}^{ab}, \\
M_{abcd} &=& \hat K_{ab} \hat K_{cd} - \hat K_{ac} \hat K_{bd}. \eea
Let us first study the expansion of $\tilde L^{ab}$. In the previous
section we calculated various elements needed to write such an
expansion. Note that \bea
\tilde L^{ab} &=& h^{mn}(L^{-1})_{mn}{}^{ab} = \frac{1}{\rho^2} \left[h^{(0)\: mn} - \frac{1}{\rho}h^{(1)\: mn} - \frac{1}{\rho^2}  \left( h^{(2)\: mn} - h^{(1)\: mp} h^{(1)}_{p}{}^{n}\right) + \ldots \right] \nn \\ & \times &\left[ (L^{-1})^{(0)}{}_{mn}{}^{ab} + \rho^{-1}(L^{-1})^{(1)}{}_{mn}{}^{ab} + \rho^{-2} (L^{-1})^{(2)}{}_{mn}{}^{ab} + \ldots \right] \nn \\
&=& \rho^{-1} \frac{h^{(0)\: ab}}{4} + \rho^{-2} \left( \frac{\s^{ab}}{2} + \s h^{(0)\: ab}\right) + \rho^{-4}\Big{[} h^{(0)\: mn} (L^{-1})^{(2)}{}_{mn}{}^{ab} \nn \\ &&- h^{(1)\: mn} (L^{-1})^{(1)}{}_{mn}{}^{ab} - \left( h^{(2)\: mn} - h^{(1)\: mp} h^{(1)}_{p}{}^{n}\right)(L^{-1})^{(0)}{}_{mn}{}^{ab}\Big{]} + \ldots \ \ . \nn \label{ltilde}\\
\eea
After some calculation we find the following expressions for the second and third elements of the second order term of equation (\ref{ltilde})
\bea
- h^{(1)\: mn} (L^{-1})^{(1)}{}_{mn}{}^{ab} &=& \rho \left( \s^2 h^{(0)\: ab} + \s \s^{ab} \right), \\
- \left( h^{(2)\: mn} - h^{(1)\: mp} h^{(1)}_{p}{}^{n}\right)(L^{-1})^{(0)}{}_{mn}{}^{ab} &= & - \rho \left( h^{(2)\: ab} - 4 \s^2 h^{(0)\: ab} - \frac{1}{4} h^{(0)\: ab}\s_c \s^c\right)~.
\eea
Adding together these expressions with (\ref{linv2}) we find that (\ref{ltilde}) simplifies to
\bea
\tilde L^{ab} &=& \rho^{-1} \frac{h^{(0)\:ab}}{4} + \rho^{-2} \left( \frac{\s^{ab}}{2} + \s h^{(0)\: ab}\right) + \frac{\rho^{-3}}{4}  \Bigg{[} 2 \hat p^{(2)\: ab}   - h^{(0) \: ab} \hat p^{(2)} - 3 h^{(2)\: ab} \nn \\ && + h^{(0)\: ab} \s_c \s^c   + \frac{65}{2} \s^2 h^{(0)\: ab} + 16 \s \s^{ab} - \frac{3}{2} \s_{cd}\s^{cd} h^{(0)\: ab} + 4 \s^{ac} \s_{c}^{b}\Bigg{]}
+ \ldots \ \ .
\label{Ltilde}
\eea
Now we want to eliminate $\hat p^{(2)\: ab}$ in favor of $\{h^{(2)\: ab}, \s\}$
and their derivatives. This is done using equation (B.10) of \cite{MMV}.
Applying the equations of motion  one can show that
\be
\hat p^{(2)}_{ab} = h^{(2)}_{ab} - h^{(0)}_{ab} \s_c \s^c + 2 \s_a \s_b - \frac{5}{4} h^{(0)}_{ab}\s^2 + \s \s_{ab} + \s_a{}{}^c \s_{cb} - \frac{1}{4} \s^{cd}\s_{cd}h^{(0)}_{ab}.
\label{phat2}
\ee
Substituting (\ref{phat2}) in (\ref{Ltilde}) and doing some algebra we find
\bea
\tilde L^{ab} &=& \rho^{-1} \frac{h^{(0)\: ab}}{4} + \rho^{-2} \left( \frac{\s^{ab}}{2} + \s h^{(0)\: ab}\right) +
\frac{\rho^{-3}}{4} \Bigg{[} \frac{99}{4} \s^2 h^{(0)\: ab} - \frac{9}{4} \s_{cd}\s^{cd} h^{(0)\: ab} \nn \\ && + 6 \s^{ac} \s_{c}{}^{b} + 18 \s \s^{ab}
- h^{(2)\: ab} + 4 \s^a \s^b - \s_c \s^c h^{(0)\: ab}\Bigg{]}
+ \ldots \ ~.
\label{Ltilde2}
\eea
The expansion of $M_{abcd}$ is relatively straight forward
\bea
M_{abcd} &=& \rho^2 (h^{(0)}_{ab} h^{(0)}_{cd} - h^{(0)}_{ac}h^{(0)}_{db}) + \rho (\hat p^{(1)}_{ab} h^{(0)}_{cd} + h^{(0)}_{ab} \hat p^{(1)}_{cd}   - \hat p^{(1)}_{ac} h^{(0)}_{bd}  - h^{(0)}_{ac} \hat p^{(1)}_{db} ) \nn \\
&+& (\hat p^{(2)}_{ab} h^{(0)}_{cd} + \hat p^{(1)}_{ab} \hat p^{(1)}_{cd}  +  h^{(0)}_{ab} \hat p^{(2)}_{cd}  - \hat p^{(2)}_{ac} h^{(0)}_{db}  - \hat p^{(1)}_{ac} \hat p^{(1)}_{db}  - h^{(0)}_{ac}\hat p^{(2)}_{db}) + \ldots  \ ~.
\eea
Using the expansion (\ref{Ltilde2}) for $\tilde L^{ab}$ and the expression (\ref{phat2}) for $\hat p^{(2)}_{ab}$ we  find
\bea
\tilde L^{ab} M_{abcd} &=&  \frac{\rho}{2} h^{(0)}_{cd} - \frac{\sigma_{cd} + 5 \sigma h^{(0)}_{cd}}{4} + \frac{1}{4\rho}
\Bigg{[} \frac{7}{2}\s_{pq}\s^{pq}h^{(0)}_{cd} -2 \s_c \s_d  -\frac{47}{2} \s^2 h^{(0)}_{cd}   + 2 h^{(2)}_{cd}\nn \\ && -  19 \s \s_{cd} -10\s_{c}{}^{p} \s_{dp}  \Bigg{]} + \ldots \ \ \label{lm1} .
\eea
The expansions (\ref{Ltilde2}) and (\ref{lm1}) are the key results of this section.

\subsection{Calculation of the Derivative Terms}
\label{section:b3} In this section we calculate the $\rho^{-1}$
expansions of the derivative terms. We start by calculating a
general expression for two derivatives acting on $\tilde L^{ab}$.
For the moment we keep all four indices free
 \bea
\mathcal{{D}}_c \mathcal{{D}}_d \tilde L^{mn} &=&
{D}^{[0]}_c {D}^{[0]}_d \tilde L^{(0) \: mn}
+  \rho^{-1}\left(
{D}^{[0]}_c {D}^{[0]}_d \tilde L^{(1) \: mn}
+ {D}^{[0]}_c {D}^{[1]}_d \tilde L^{(0) \: mn}
\right)  \nn \\ & &
+ \rho^{-2}\Big{(}
{D}^{[0]}_c {D}^{[0]}_d \tilde L^{(2) \: mn}
+  {D}^{[0]}_c {D}^{[1]}_d \tilde L^{(1) \: mn}
+ {D}^{[0]}_c {D}^{[2]}_d \tilde L^{(0) \: mn}
\nn \\ &&
+ {D}^{[1]}_c {D}^{[1]}_d \tilde L^{(0) \: mn}
+ {D}^{[1]}_c {D}^{[0]}_d \tilde L^{(1) \: mn}
\Big{)} + \ldots \ \ .
\label{deri}
\eea
In expression (\ref{deri}) ${D}^{[0]}$ is simply $D$ -- the torsion free covariant derivative compatible with the unit hyperboloid metric $h^{(0)}_{ab}$. ${D}^{[1]}$ and ${D}^{[2]}$ respectively denote the first order and the second order corrections to the covariant derivative $\mathcal{D}$ away from $D$.
Calculating various terms in equation (\ref{deri}) requires some amount of work. Using equations (\ref{c1}) and (\ref{c2}) of appendix \ref{collection} we find the following results:
\begin{enumerate}
\item
\bea
{D}^{[0]}_c {D}^{[0]}_d \tilde L^{(1) \: mn} &=& \frac{1}{2} \s^{mn}{}_{dc} + \s_{dc}h^{(0)\:mn}
\eea
\item
\bea
{D}^{[0]}_c {D}^{[1]}_d \tilde L^{(0) \: mn} &=& - \frac{1}{2} h^{(0)\: mn} \s_{cd}
\eea
\item
\bea
{D}^{[0]}_c {D}^{[0]}_d \tilde L^{(2) \: mn} &=& \frac{1}{4} \Bigg{[} \frac{99}{2} \s_c \s_d h^{(0)\: mn} + \frac{99}{2} \s \s_{cd}h^{(0)\: mn} \nn \\ & & - \frac{9}{2}\s_{pq}\s^{pq}{}_{dc} h^{(0)\: mn} - \frac{9}{2} \s_{pqc}\s^{pq}{}_{d} h^{(0)\: mn} \nn \\
& &+ 6 \s^{mp}{}_{dc} \s_p{}^{n}  +6 \s^{mp}{}_{d} \s_p{}^{n}{}_{c} + 6 \s^{mp}{}_{c} \s_p{}^{n}{}_{d}+ 6 \s^{mp} \s_p{}^{n}{}_{dc}  \nn \\ & & + 18\s_{dc}\s^{mn} + 18 \s_d \s^{mn}{}_{c} + 18 \s_c \s^{mn}{}_{d} + 18 \s \s^{mn}{}_{dc}  \nn \\ & & - h^{(2)\: mn}{}_{;dc}\nn \\  &  & + 4 \s^m{}_{dc} \s^n + 4 \s^m{}_{d} \s^n{}_{c} + 4 \s^m{}_{c} \s^n{}_{d} + 4\s^m\s^{n}{}_{dc} \nn \\ &&- 2 \s_{pdc} \s^{p} h^{(0)\: mn} -   2\s_{pd}\s^p{}_{c} h^{(0) \: mn} \Bigg{]}
\eea
\item
\bea
{D}^{[0]}_c {D}^{[1]}_d \tilde L^{(1) \: mn} &=& \Bigg{[} - \s_{ec} \s^{e(n}h^{(0)}_{d}{}^{m)} - \s_e \s^{e(n}{}_{c} h^{(0)}_d{}^{m)} - \s_{dc} \s^{mn} - \s_d \s^{mn}{}_{c} \nn \\
& & - 2\s_{dc} \s h^{(0)\: mn} - 2 \s_d \s_c h^{(0)\: mn} + \s^{(m}{}_{c} \s_{d}{}^{n)} + \s^{(m}\s_{d}{}^{n)}{}_{c} \Bigg{]}
\eea
\item
\bea
{D}^{[0]}_c {D}^{[2]}_d \tilde L^{(0) \: mn} &=& -  \s_c \s_d h^{(0)\: mn} -  \s \s_{cd} h^{(0)\: mn} + \frac{1}{4} h^{(2)\: mn}{}_{;dc}
\eea
\item
\bea
{D}^{[1]}_c {D}^{[1]}_d \tilde L^{(0) \: mn} &=& \frac{1}{2} \s^e \s_e h^{(0)\: mn} h^{(0)}_{cd}
\eea
\item
\bea
{D}^{[1]}_c {D}^{[0]}_d \tilde L^{(1) \: mn} &=& \frac{1}{2} \Bigg{[} 2 \s_{[d} \s^{mn}{}_{c]}  - h^{(0)}_{cd} \s^e \s^{mn}{}_{e} - 2 \s_e \s^{e(n}{}_{d} h^{(0)\: m)}{}_{c} + 2 \s_c{}^{(n}{}_{d}\s^{m)} \nn \\ &-& 2 \s^e \s_e h^{(0)\: mn} h^{(0)}_{cd}\Bigg{]}.
\eea
\end{enumerate}
Various contractions of the above expressions give the required three final terms in (\ref{main}).


\subsubsection*{Term 1}
\begin{equation}
h_{ab} \mathcal{{D}}_{m}\mathcal{{D}}_{n}\tilde L^{mn} = \frac{1}{8 \, \rho} \,\left[ 45 \s^2  - 29 \s_m \s^m + 5 \s_{mn}\s^{mn} + 3 \s_{mnp}\s^{mnp}\right]h^{(0)}_{ab} + \ldots ~.
\end{equation}
This result uses the following two identities from the appendix \ref{collection}
\bea
\s^{mp}{}_{nm} \s_{p}{}^{n} &=& 2 \s_{mn}\s^{mn} - 9 \s^2 \, ,  \\
\s^{mp}{}_{n}\s_{p}{}^{n}{}_{m} &=& \s_{mnp}\s^{mnp} - 2 \s_m \s^m \, .
\eea

\subsubsection*{Term 2}
\begin{eqnarray}
\mathcal{{D}}^2 \tilde L_{ab} &=& \frac{3}{2} \, \left( \sigma h^{(0)}_{ab} + \sigma_{ab}\right)
+ \frac{1}{4\rho}\Bigg{[}- \frac{9}{2} \,\s_{mnp}\s^{mnp} h^{(0)}_{ab} + \frac{51}{2} \,\s_m \s^m h^{(0)}_{ab} +
\frac{129}{2}\, \s^2 h^{(0)}_{ab} - 8 \,\s_a \s_b \nn \\ & & \quad - \frac{31}{2} \, \s_{mn}\s^{nm} h^{(0)}_{ab}
+ 44 \,\s_{am}\s^{m}{}_{b} + 72 \,\s \s_{ab} + 12 \,\s_{a}{}^{mn} \s_{bmn} + 18 \,\s^{e}\s_{abe}  \Bigg{]} + \ldots ~ .
\end{eqnarray}
The calculation of this term relies on the following identity from appendix \ref{collection}
\bea
\s_e\s^{nem} &=& \s_e \s^{mne} - \s^m \s^n + \s^e \s_e h^{(0)}_{mn}.
\eea



\subsubsection*{Term 3}
The calculation of this term is more involved than the previous two terms. We find
\bea
- 2 \mathcal{{D}}_{n}\mathcal{{D}}_{(b}\tilde L_{a)}{}^{n}  &=& - 3 \left( \sigma_{ab} + \sigma h^{(0)}_{ab} \right) + \frac{1}{\rho} \Bigg{[} \frac{5}{4}\,\s_a\s_b - \frac{51}{4}\,\s \s_{ab} - 11\,\s_{ap}\s_b{}^{p} + \s_{apq}\s^q + 3 \,\s_p \s^p h^{(0)}_{ab} \nn \\
&& - 21 \,\s^2 h^{(0)}_{ab}+ 4 \,\s_{mn}\s^{mn}h^{(0)}_{ab}  -\frac{3}{4}\,\s^{pq}\s_{pq(ab)}  - \frac{3}{4}\,\s^{pq}{}_{(a}\s_{|pq|b)} \Bigg{]} + \ldots \ \ \ .
\eea
The identities from appendix \ref{collection} that are needed for this calculation are
\bea
\s_{p(ab)q}\s^{pq} &=& \s_{pq(ab)}\s^{pq} - 6 \,\s \s_{ab} - h^{(0)}_{ab} \s_{pq}\s^{pq} - \s_{ap}\s^{p}{}_{b} \ \ , \\
\s_{(a}{}^{pq}\s_{|pq|b)} &=& \s_{pq(a}\s^{pq}{}_{b)}  - 3 \, \s_a \s_b - \s^q\s_{q(ab)}  \\
&=& \s_{pq(a}\s^{pq}{}_{b)} - 2 \, \s_a \s_b - \s^q \s_{abq} - h^{(0)}_{ab}\s_p \s^{p} \ \ , \\
\s_{(a}{}^{n}{}_{b)n} &=& 2 \,\s_{ab} + 3 \,\s h^{(0)}_{ab} ~ .
\eea

\subsection{A Final expression for $\Delta_{ab}$}
\label{section:b4}
In this section we collect the expansions performed in the previous sections and give a final expression for $\Delta_{ab}$. Recall that
\be
\Delta_{ab}= \hat{K}_{ab}-2\tilde{L}^{cd}M_{abcd} + \mathcal{D}^{2}\tilde{L}_{ab}+h_{ab}\mathcal{D}_{k}\mathcal{D}_{l}%
\tilde{L}^{kl}-2 \mathcal{D}_{k} \mathcal{D}_{(a}\tilde{L}^{k}{}_{b)}.
\ee
Various terms in $\Delta_{ab}$ admit the following expansions:
\begin{enumerate}
\item
\bea
\hat  K_{ab} &=& \rho \,h^{(0)}_{ab} + \s_{ab} - \s \,h^{(0)}_{ab} + \frac{1}{\rho}\,\Bigg{[}h^{(2)}_{ab} - h^{(0)}_{ab} \s_c \s^c + 2 \,\s_a \s_b - \frac{5}{4}\, h^{(0)}_{ab}\s^2 + \s \s_{ab} + \s_a{}{}^c \s_{cb} \nn \\ &&  - \frac{1}{4} \,\s^{cd}\s_{cd} \, h^{(0)}_{ab}\Bigg{]} + \ldots ~ ,
\label{p2}
\eea
\item
\bea
- \tilde L^{ab} M_{abcd} &=&  - \frac{\rho}{2} \, h^{(0)}_{cd} + \frac{1}{4}\left(\sigma_{cd} + 5 \,\sigma\, h^{(0)}_{cd}\right) - \frac{1}{4 \, \rho} \,
\Bigg{[} \frac{7}{2}\,\s_{pq}\s^{pq} \, h^{(0)}_{cd} -2 \,\s_c \s_d  -\frac{47}{2} \,\s^2 h^{(0)}_{cd}   +
2 \, h^{(2)}_{cd}\nn \\ && -  19 \, \s \, \s_{cd} -10 \, \s_{c}{}^{p} \s_{dp}  \Bigg{]} + \ldots \ \ \label{lm} ,
\eea
\item
\bea
h_{ab} \mathcal{{D}}_{k}\mathcal{{D}}_{l}\tilde L^{kl} &=& \frac{1}{8 \, \rho} \, \left[ 45 \, \s^2  - 29 \, \s_m \s^m +
5 \, \s_{mn}\s^{mn} + 3 \,\s_{mnp}\s^{mnp} \right]\,h^{(0)}_{ab} + \ldots \ \ ,
\eea
\item
\bea
\mathcal{{D}}^2 \tilde L_{ab} &=& \frac{3}{2} \, \left( \sigma h^{(0)}_{ab} + \sigma_{ab}\right)
+ \frac{1}{4\,\rho}\,\Bigg{[}- \frac{9}{2} \,\s_{mnp}\s^{mnp} \,h^{(0)}_{ab} + \frac{51}{2} \,\s_m \s^m \,h^{(0)}_{ab} +
\frac{129}{2}\, \s^2 \,h^{(0)}_{ab} - 8 \,\s_a \s_b \nn \\ && -\frac{31}{2} \,\s_{mn}\s^{nm} \,h^{(0)}_{ab}
+ 44 \,\s_{am}\s^{m}{}_{b} + 72 \,\s \s_{ab} + 12 \,\s_{a}{}^{mn} \s_{bmn} + 18 \,\s^{e}\s_{abe}  \Bigg{]} + \ldots \ \  ,
\eea
\item
\bea
- 2 \,\mathcal{{D}}_{n}\mathcal{{D}}_{(b}\tilde L_{a)}{}^{n}  &=& - 3 \left( \sigma_{ab} + \sigma \,h^{(0)}_{ab} \right)+ \frac{1}{\rho} \,\Bigg{[} \frac{5}{4}\,\s_a\s_b - \frac{51}{4}\,\s \s_{ab} - 11\,\s_{ap}\s_b{}^{p} + \s_{apq} \s^q +
3 \, \s_p \s^p \, h^{(0)}_{ab} \nn \\
&& - 21 \,\s^2 \,h^{(0)}_{ab}+ 4 \,\s_{mn}\s^{mn}\,h^{(0)}_{ab}  -\frac{3}{4}\,\s^{pq}\s_{pq(ab)}  - \frac{3}{4}\,\s^{pq}{}_{(a}\s_{|pq|b)} \Bigg{]} + \ldots \ \ \ .
\eea
\end{enumerate}
Adding these terms we find
\bea
\Delta_{ab} &=:& \frac{1}{\rho}\,\Delta^{(2)}_{ab} + \mathcal{O} \left(\frac{1}{\rho^2}\right) \nn \\
&=&  \frac{1}{4\,\rho}\,
\Bigg{[}
9 \,\s_c \s^c \,h^{(0)}_{ab} - 29 \,\s_a \s_b + 63 \,\s\s_{ab} + 24 \,\s_{ap}\s_{b}{}^{p} - 5 \, \s_{cd}\s^{cd}\,h^{(0)}_{ab} + 45 \,\s^2 \,h^{(0)}_{ab} \nn \\ &-& 3\,\s_{mnp}\s^{mnp}\,h^{(0)}_{ab} + 9 \,\s_{pq(a}\s^{pq}{}_{b)} - 3 \, \s^{pq}\s_{pq(ab)} - 2\, \s^e\s_{e(ab)}
\Bigg{]} + \ldots \ \  \label{final}.
 \eea
The expression (\ref{final}) is the key result of this appendix. Notice that all of the $h^{_{(2)}}_{ab}$ terms have canceled, so that $\Delta_{ab}^{_{(2)}}$ is determined by $\sigma$ and $h_{ab}^{_{(0)}}$ .

\setcounter{equation}{0}

\section{Stress Tensor Calculation $d>4$}
\label{astdg4}

This appendix collects a number of calculations for asymptotically flat spacetimes in $d>4$ dimensions. In section \ref{section:a0} we perform the asymptotic expansion of the field equations, then calculate the counterterm $\hat{K}_{ab}$
to the requisite order section \ref{section:a1}. These results are used in sections \ref{section:a2} and \ref{section:a3} to obtain an asymptotic expression for $\Delta_{ab}$. Remarkably, $\Delta_{ab}$ vanishes identically up to terms that vanish too rapidly (as $\rho \to \infty$) to contribute to the conserved charges. We close in section \ref{section:a4} by presenting the complete expression for the boundary stress-energy. For convenience we set $n=d-3$ in parts of this appendix.

\subsection{Asymptotic Expansion: $d>4$}
\label{section:a0}

We now perform an asymptotic expansion of the $d > 4$ field
equations to obtain the results (\ref{Rab0}-\ref{ee13b}), analogous
to those of \cite{B,BS} for $d=4$.  See also \cite{skenAF} for
similar results.

The expansion of the extrinsic curvature for $d>4$ is
\be P_{ab} =\rho
^{-1}K_{ab}=h_{ab}^{^{(0)}}-\frac{1}{\rho ^{d-3}}\left( \sigma
h_{ab}^{^{(0)}}+ \frac{d-5}{2}h_{ab}^{^{(1)}}\right)
-\frac{d-4}{2}\frac{h_{ab}^{^{(2)}}}{\rho ^{d-2}}  +
\mathcal{O}\left(\frac{1}{\rho^{d-1}}\right)~,\label{e11e}
\ee%
where for convenience we define
$P_{ab}:=\rho ^{-1}K_{ab}$. The trace of the extrinsic curvature is
\begin{equation}
K= \frac{\left( d-1\right)}{\rho} -%
\frac{1}{\rho ^{d-2}}\left( \left( d-1\right) \sigma +\frac{d-3}{2}%
h^{^{(1)}}\right) +\frac{\left( d-2\right) }{2\,\rho ^{d-1}}h^{^{(2)}}
+ \mathcal{O}\left(\frac{1}{\rho^{d}}\right) \label{ee8}
\end{equation}%
where \ $h^{{(1)}}=h^{{(0)}ab}h_{ab}^{{(1)}}$ and $%
h^{{(2)}}=h^{{(0)}ab}h_{ab}^{{(2)}}$.  The acceleration is \be a_{b } =
-N\mathcal{D}_{b}\sigma  =  \left(0, -\frac{1}{\rho
^{d-3}}D_{b}\sigma\right) \label{ee4a} \ee to the appropriate order.
As in the main text, $D_a$ is the torsion-free covariant derivative
on ${\mathcal H}$ compatible with $h^{_{(0)}}_{ab}$.

Inserting these expansions into the Einstein equation (\ref{ee1})
yields
 \begin{eqnarray}
0 &=&\mathcal{R}_{ab}^{{(0)}}-\left( d-2\right) h_{ab}^{{(0)}}
\nonumber
\\
&&+\frac{1}{\rho ^{d-3}}\left[
\mathcal{R}_{ab}^{{(1)}}-D_{a}D_{b}\sigma
+\left(d-1\right) \sigma h_{ab}^{{(0)}}+\frac{d-3}{2}%
h^{{(1)}}h_{ab}^{{(0)}}-\frac{\left(d-1\right)
}{2}h_{ab}^{{(1)}}\right]
\label{ee10b2} \\ \nn
&&+\frac{1}{\rho ^{d-2}}\left[ \mathcal{R}_{ab}^{{(2)}}-\left(
d-2\right) h_{ab}^{{(2)}}+\frac{\left(d-2\right)
}{2}h^{{(2)}}h_{ab}^{{(0)}}\right] +
\mathcal{O}\left(\frac{1}{\rho^{d-1}}\right)~.
 \end{eqnarray}%
The leading term in this equation contains $\mathcal{R}_{ab}^{_{(0)}}$, the Ricci tensor for the metric $h_{ab}^{_{(0)}}$. The sub-leading terms depend on the tensors appearing in the asymptotic expansion of $\mathcal{R}_{ab}$, defined as
\begin{equation}\label{Rabn}
\mathcal{R}_{ab}^{(m)} = \frac{1}{2}\,\left(D^{c} D_{a} h_{cb}^{(m)} + D^{c} D_{b} h_{ac}^{(m)} - D^{c} D_{c} h_{ab}^{(m)} - D_{a} D_{b} h^{(m)} \right)
\end{equation}
for $m=1,2$. The expansion of equation (\ref{ee2}) gives
\bea 0 & =& \frac{1}{\rho
  ^{d-3}}\left[ \left( d-2\right) D^{a}\sigma -\frac{d-3}{2}\left(
  D_{b}h^{{(1)}ab}-D^{a}h^{{(1)}}\right) \right] \label{ee112} \\ & &  -\frac{\left(d-2\right) }{2%
  }\frac{1}{\rho ^{d-2}}\left[
  D_{b}h^{{(2)}ab}-D^{a}h^{{(2)}}\right]+
  \mathcal{O}\left(\frac{1}{\rho^{\,d-1}}\right) \nn,
\eea
where indices are raised and with respect to the metric $h_{ab}^{{(0)}}.$ The remaining equation (\ref{ee3}) can be written as
\begin{equation}
\mathcal{R}-K^{2} + K^{ab} K_{ab} = 0
\end{equation}
which yields
\begin{eqnarray}
0 &=&\mathcal{R}^{{(0)}}-\left( d-2\right) \left( d-1\right)  \nonumber \\
&&+\frac{1}{\rho ^{d-3}}\left[ \mathcal{R}^{{(1)}}+2\left(
d-2\right) \left(
d-1\right) \sigma +\left( d-2\right)\left( d-4\right) h^{{(1)}}\right]  \nonumber \\
&&+\frac{1}{\rho ^{d-2}}\left[ \mathcal{R}^{{(2)}}+\left(
d-2\right)\left( d-3\right) h^{{(2)}}\right] +
\mathcal{O}\left(\frac{1}{\rho^{d-1}}\right)~. \label{ee12}
\end{eqnarray}%
Note that $\mathcal{R}^{(m)}$ is defined as the trace of (\ref{Rabn})
\begin{equation}
  \mathcal{R}^{(m)} = h^{(0)ab}\,\mathcal{R}_{ab}^{(m)} ~.
\end{equation}
This is \emph{not} the same as the term at order $m$ in the
asymptotic expansion of $\mathcal{R}$.

To leading order the Einstein equations become
\begin{eqnarray}
\mathcal{R}_{ab}^{{(0)}}-(d-2) \, h_{ab}^{{(0)}} &=& 0 \label{e12a} ~.
\end{eqnarray}
The subleading equations are obtained from (\ref{ee10b2}), (\ref{ee112}) and (\ref{ee12}) by
setting the coefficients of the various powers of     $\rho^{-1}$ to
zero.  Note that these equations are not all independent, and their
self-consistency imposes additional constraints on the asymptotic
expansion.  For example, taking the trace of (\ref{ee10b2}) with
respect to $h_{ab}^{_{(0)}}$ implies that
\begin{eqnarray}
D^{2}\sigma +(d-3)\left(d-1\right) \sigma +\frac{(d-3)\,\left( d-4\right) }{2}%
h^{{(1)}} &=&0,   \label{ee13a2} \\
{\rm and} \ \ \ h^{{(2)}} &=&0,  \label{ee13b2}
\end{eqnarray}%
where we have used (\ref{ee112}) and (\ref{Rabn}).

The leading term in the Einstein equations has different implications for asymptotically flat boundary conditions in $d>4$ and $d=4$. When $d=4$, spatial infinity is a three dimensional surface and so the equation $R_{ab}^{_{(0)}} = 2\,h_{ab}^{_{(0)}}$ implies that the metric on $\mathcal{H}$ is locally that of the unit hyperboloid. For $d>4$ the metric on $\mathcal{H}$ may have a non-vanishing Weyl tensor, so $h_{ab}^{_{(0)}}$ contains degrees of freedom that are not determined by (\ref{Rab0}). For simplicity we still refer to $\mathcal{H}$ as the unit hyperboloid, but in fact we only require the condition (\ref{Rab0}). All of our results in $d>4$ apply for metrics $h_{ab}^{_{(0)}}$ that have arbitrary Weyl curvature.

\subsection{Extrinsic Curvature Counterterm}
\label{section:a1}

We now compute the counter-term ${\hat K}_{ij}$ to the required
order. We employ the asymptotic field equations
(\ref{ee10b2}-\ref{ee12}), expanded in powers of
$\rho^{-1}$, to simplify results as needed.

To leading asymptotic order the Einstein equations imply that
\be
\hat{K}_{ij}^{{(0)}}= {\rho }\,h_{ij}^{{(0)}}
\ee
using equations (\ref{ee10b2}) and (\ref{Khat}). Expanding this latter equation we
have
\begin{equation}
\hat{P}\hat{P}_{ab}-h^{cd}\hat{P}_{ac}\hat{P}_{db}=\mathcal{R}%
_{ab}^{{(0)}}+\frac{\mathcal{R}_{ab}^{{(1)}}}{\rho ^{n}}+\frac{\mathcal{R}%
_{ab}^{{(2)}}}{\rho ^{n+1}}   + \mathcal{O}\left(
\frac{1}{\rho^{n+2}}\right)\label{e11f}
\end{equation}%
where \ $\hat{P}_{ab}=\rho ^{-1}\hat{K}_{ab}$ admits the expansion
\begin{equation}
\hat{P}_{ab}=h_{ab}^{{(0)}}+\frac{1}{\rho ^{n}}\hat{P}_{ab}^{{(1)}}+\frac{1%
}{\rho ^{n+1}}\hat{P}_{ab}^{{(2)}} +\mathcal{O}\left(
\frac{1}{\rho^{n+2}}\right) ~. \label{ec1}
\end{equation}%
It is straightforward to solve iteratively for $\hat{P}_{ab}^{(I)}$.
This gives
\begin{eqnarray}
\hat{P}_{ab}^{{(1)}} &=&\frac{1}{n}\left[ \mathcal{R}_{ab}^{{(1)}}-\frac{1%
}{2\left( n+1\right) }\mathcal{R}^{{(1)}}h_{ab}^{{(0)}}-h_{ab}^{{(1)}}+%
\frac{1}{2}h^{{(1)}}h_{ab}^{{(0)}}\right]  \label{ec2} \\
\hat{P}_{ab}^{{(2)}} &=&\frac{1}{n}\left[ \mathcal{R}_{ab}^{{(2)}}-\frac{1%
}{2\left( n+1\right) }\mathcal{R}^{{(2)}}h_{ab}^{{(0)}}-h_{ab}^{{(2)}}+%
\frac{1}{2}h^{{(2)}}h_{ab}^{{(0)}}\right]  \label{ec3}
\end{eqnarray}
where indices are raised/lowered and traces taken with respect to
$h^{_{(0)}}_{ab}$.
When the Einstein equations are satisfied these expressions simplify to
\begin{eqnarray}
\hat{P}_{ab}^{{(1)}} &=&\frac{1}{n}D_{a}D_{b}\sigma +\frac{1}{2}%
h_{ab}^{{(1)}}\qquad  \hat{P}^{{(1)}}=\frac{1}{n}%
D^{2}\sigma +\frac{1}{2}h^{{(1)}}  \label{ec4a} \\
\hat{P}_{ab}^{{(2)}} &=&h_{ab}^{{(2)}}
\qquad  \hat{P}^{{(2)}}=h^{{(2)}} = 0
\label{ec4b}
\end{eqnarray}
where we recall that $h^{{(2)}}=0$ by the trace (\ref{ee13b2}) of the second order term in the Einstein equations.

\subsection{Computation of  the Inverse of $L_{ab}{}^{cd}$}
\label{section:a2}

Recall that the inverse of $L_{ab}{}^{cd}$ is defined by \be
(L^{-1})_{ij}{}^{kl} L_{kl}{}^{mn} = \delta_{i}^{m} \delta_{j}^{n}.
\label{L-inv} \ee Following \cite{MM} we introduce the expansions
\bea L &=& L^{(0)} + \rho^{-(d-3) }L^{(1)} + \rho^{-(d-2)}L^{(2)} + \ldots
\\ \mbox{ and } \left(L^{-1}\right) &=& \left(L^{-1}\right)^{(0)} +
\rho^{-(d-3)}\left(L^{-1}\right)^{(1)} +
\rho^{-(d-2)}\left(L^{-1}\right)^{(2)} +\ldots \, . \label{a10} \eea
 Using the definition of $L^{-1}$ we find \be
\left(L^{-1}\right)^{(1)}{}_{ij}{}^{pq} = -
\left(L^{-1}\right)^{(0)}{}_{ij}{}^{kl}
\left(L^{(1)}\right)_{kl}{}^{mn}
\left(L^{-1}\right)^{(0)}{}_{mn}{}^{pq}, \label{L1} \ee and \be
\left(L^{-1}\right)^{(2)}{}_{ij}{}^{pq} = -
\left(L^{-1}\right)^{(0)}{}_{ij}{}^{kl}
\left(L^{(2)}\right)_{kl}{}^{mn}
\left(L^{-1}\right)^{(0)}{}_{mn}{}^{pq}. \label{L2} \ee

Defining $Q_{ij}{}^{kl}=\rho ^{-1}L_{ij}{}^{kl}$ we have from
(\ref{e6}) that
\begin{equation}
Q_{ij}{}^{kl}=h^{kl}\hat{P}_{ij}+\delta _{i}^{(k}\delta _{j}^{l)}\hat{P}%
-\delta _{i}^{(k}\hat{P}_{j}^{l)}-\delta _{j}^{(k}\hat{P}_{i}^{l)}
\label{l1}
\end{equation}%
and upon expansion we obtain
\begin{eqnarray}
Q_{ij}^{^{(0)}}{}^{kl} &=&h_{ij}^{^{(0)}}h^{^{(0)}kl}+n\delta
_{i}^{(k}\delta _{j}^{l)}  \label{l2a} \\
Q_{ij}^{^{(1)}}{}^{kl} &=&\hat{P}%
_{ij}^{^{(1)}}h^{^{(0)}kl}-h_{ij}^{^{(0)}}h^{^{(1)}kl}+\delta
_{i}^{(k}\delta _{j}^{l)}\left( \hat{P}^{^{(1)}}-h^{^{(1)}}\right)
-\delta _{i}^{(k}\left(
\hat{P}_{j}^{^{(1)}l)}-h_{j}^{^{(1)}l)}\right) \nn \\ & & -\delta
_{j}^{(k}\left( \hat{P}_{i}^{^{(1)}l)}-h_{i}^{^{(1)}l)}\right) \label{l2b}\\
Q_{ij}^{^{(2)}}{}^{kl} &=&\hat{P}%
_{ij}^{^{(2)}}h^{^{(0)}kl}-h_{ij}^{^{(0)}}h^{^{(2)}kl}+\delta
_{i}^{(k}\delta _{j}^{l)}\left( \hat{P}^{^{(2)}}-h^{^{(2)}}\right)
-\delta _{i}^{(k}\left(
\hat{P}_{j}^{^{(2)}l)}-h_{j}^{^{(2)}l)}\right) \nn \\ & & -\delta
_{j}^{(k}\left( \hat{P}_{i}^{^{(2)}l)}-h_{i}^{^{(2)}l)}\right)
\label{l2c}
\end{eqnarray}%
From (\ref{L1}, \ref{L2}) we can compute the expansion of $\left(
Q^{-1}\right) _{ij}{}^{^{(0)}kl}$. However the quantity that we
really need is $\tilde{Q}^{ab}=\frac{1}{\rho
^{2}}h^{rs}Q_{rs}{}^{ab}=\frac{1}{\rho ^{3}}h^{rs}L_{rs}{}^{ab}=\rho ^{-1}%
\tilde{L}^{ab}$, which upon expansion is%
\begin{eqnarray}
\tilde{Q}{}^{^{(0)}ab} &=&\frac{1}{2\left( n+1\right) }h^{^{(0)}ab},
\label{l4a} \\
\tilde{Q}{}^{^{(1)}ab} &=&\frac{1}{n\left( n+1\right) }\left[ \hat{P}%
^{^{(1)}ab}-\frac{1}{2}\hat{P}^{^{(1)}}h^{^{(0)}ab}-\frac{n+2}{2}%
h^{^{(1)}ab}+\frac{1}{2}h^{^{(1)}}h^{^{(0)}ab}\right],   \label{l4b} \\
\tilde{Q}{}^{^{(2)}ab} &=&\frac{1}{n\left( n+1\right) }\left[ \hat{P}%
^{^{(2)}ab}-\frac{1}{2}\hat{P}^{^{(2)}}h^{^{(0)}ab}-\frac{n+2}{2}%
h^{^{(2)}ab}+\frac{1}{2}h^{^{(2)}}h^{^{(0)}ab}\right],   \label{l4c}
\end{eqnarray}%
or, using the Einstein equations (i.e., using (\ref{ec4a}) and
(\ref{ec4b})):
\begin{eqnarray}
\tilde{Q}{}^{^{(0)}ab} &=&\frac{1}{2\left( n+1\right) }h^{^{(0)}ab},
\label{l5a} \\
\tilde{Q}{}^{^{(1)}ab} &=&\frac{1}{2n\left( n+1\right) }\left[ \frac{2}{n}%
D^{a}D^{b}\sigma -\left( n+1\right) h^{^{(1)}ab}+\left( n+2\right)
\sigma
h^{^{(0)}ab}+\frac{n}{2}h^{^{(1)}}h^{^{(0)}ab}\right],   \label{l5b} \\
\tilde{Q}{}^{^{(2)}ab} &=&-\frac{1}{2\left( n+1\right) }h^{^{(2)}ab}
~. \label{l5c}
\end{eqnarray}

\subsection{Finding the Full Variation}
\label{section:a3}

We now need to use (\ref{main})  to compute $\Delta_{ab}$
\begin{equation}
\Delta^{ab} = \hat{K}^{ab}-2\tilde{L}^{cd}\left( \hat{K}_{cd}\hat{%
K}^{ab}-\hat{K}_{c}^{\phantom{c}a}\hat{K}_{d}^{\phantom{c}b}\right) +\mathcal{D%
}^{2}\tilde{L}^{ab}+h^{ab}\mathcal{D}_{k}\mathcal{D}_{l}\tilde{L}^{kl}-%
\mathcal{D}_{k}\left( \mathcal{D}^{a}\tilde{L}^{kb}+\mathcal{D}^{b}\tilde{L}%
^{ka}\right)  \label{f1}
\end{equation}%
which is more easily done in pieces.

We begin with the first two terms. We have
\begin{equation}
\hat{K}^{ab}-2\tilde{L}^{cd}\left( \hat{K}_{cd}\hat{K}^{ab}-\hat{K}_{c}^{%
\phantom{c}a}\hat{K}_{d}^{\phantom{c}b}\right) =\frac{1}{\rho ^{3}}\left( \hat{%
P}^{ab}-2\tilde{Q}^{cd}\left(
\hat{P}_{cd}\hat{P}^{ab}-\hat{P}_{c}^{\phantom{c}a}\hat{P}_{d}^{\phantom{c}b}\right)
\right)   \label{f2}
\end{equation}%
The zeroth order term is \be
\left[ \hat{P}^{ab}-2\tilde{Q}^{cd}\left( \hat{P}_{cd}\hat{P}^{ab}-\hat{P}%
_{c}^{\phantom{c}a}\hat{P}_{d}^{\phantom{c}b}\right) \right]
^{^{(0)}} =h^{^{(0)}ab}-\frac{2}{2\left( n+1\right)
}h^{^{(0)}cd}\left( h_{cd}^{^{(0)}}h^{^{(0)}ab}-\delta
_{c}^{\phantom{c}a}\delta _{d}^{b}\right)  = 0 \ee as expected.  The
first order term is
\begin{eqnarray}
&&\left[ \hat{P}^{ab}-2\tilde{Q}^{cd}\left( \hat{P}_{cd}\hat{P}^{ab}-\hat{P}%
_{c}^{\phantom{c}a}\hat{P}_{d}^{\phantom{c}b}\right) \right]
^{^{(1)}}
\nonumber \\
&=&\hat{P}^{^{(1)}ab}-2h^{^{(1)}ab}-\frac{h^{^{(0)}cd}}{\left( n+1\right) }%
\left( h_{cd}^{^{(0)}}\left( \hat{P}^{^{(1)}ab}-2h^{^{(1)}ab}\right) +\hat{P}%
_{cd}^{^{(1)}}h^{^{(0)}ab}\right) +\frac{2\left( \hat{P}%
^{^{(1)}ab}-h^{^{(1)}ab}\right) }{\left( n+1\right) }  \nonumber \\
&&-\frac{2}{n\left( n+1\right) }\left[ \hat{P}^{^{(1)}cd}-\frac{1}{2}\hat{P}%
^{^{(1)}}h^{^{(0)}cd}-\frac{n+2}{2}h^{^{(1)}cd}+\frac{1}{2}%
h^{^{(1)}}h^{^{(0)}cd}\right] \left(
h_{cd}^{^{(0)}}h^{^{(0)}ab}-\delta
_{c}^{\phantom{c}a}\delta _{d}^{\phantom{c}b}\right)   \nonumber \\
&=&\frac{1}{n\left( n+1\right) }\left[ \frac{n+2}{n}D^{a}D^{b}\sigma -\frac{%
n+2}{2}h^{^{(1)}ab}+\left( \left( n+2\right) \sigma +\frac{n}{2}%
h^{^{(1)}}\right) h^{^{(0)}ab}\right].   \label{f4}
\end{eqnarray}%
The second order term has a similar derivation and so \bea
& & \left[ \hat{P}^{ab}-2\tilde{Q}^{cd}\left( \hat{P}_{cd}\hat{P}^{ab}-\hat{P}%
_{c}^{\phantom{c}a}\hat{P}_{d}^{\phantom{c}b}\right) \right] ^{^{(2)}} \nn \\ & =& \frac{1%
}{n\left( n+1\right) } \left[ \left( n+2\right) \left( \hat{P}%
^{^{(2)}ab}-h^{^{(2)}ab}\right) -\left(
\hat{P}^{^{(2)}}-h^{^{(2)}}\right) h^{^{(0)}ab}\right] \nn \\   &=&
0  \label{f5} \eea where the last equality follows from
(\ref{ec4b}).

We now turn to the last four terms in equation (\ref{f1}).  Using
$\tilde{L}^{ab} = \rho \tilde{Q}^{ab}$ we obtain
 for the zeroth order term
 \begin{eqnarray}
&&\left[ \mathcal{D}^{2}\tilde{Q}^{ab}+h^{ab}\mathcal{D}_{k}\mathcal{D}_{l}%
\tilde{Q}^{kl}-\mathcal{D}_{k}\left( \mathcal{D}^{a}\tilde{Q}^{kb}+\mathcal{D%
}^{b}\tilde{Q}^{ka}\right) \right] ^{^{(0)}}  \nonumber \\
&=&\frac{1}{2\left( n+1\right) }\left(
D^{2}h^{^{(0)}ab}+h_{k}^{ab}D_{k}D_{l}h^{^{(0)}kl}-D_{k}\left(
D^{a}h^{^{(0)}kb}+D^{b}h^{^{(0)}ka}\right) \right)   \nonumber \\
&=&0  \label{f7}
\end{eqnarray}
again as expected.

For the first order term we rewrite equation (\ref{l5b}) as \be
\tilde{Q}{}^{^{(1)}ab} =X^{ab}-\frac{1}{2\left( n+1\right)
}h^{^{(1)}ab}  \label{f8} \ee where \be
X_{ab} = \frac{1}{n\left( n+1\right) }\left[ \frac{1}{n}D^{a}D^{b}\sigma -\frac{1}{%
2}h^{^{(1)}ab}+\frac{1}{2}\left( \left( n+2\right) \sigma +\frac{n}{2}%
h^{^{(1)}}\right) h^{^{(0)}ab}\right]. \ee
Making this separation yields the following advantage%
\begin{eqnarray}
\left[ \mathcal{D}_{a}\mathcal{D}_{b}\tilde{Q}^{rs}\right] ^{^{(1)}}
&=&D_{a}D_{b}\tilde{Q}{}^{^{(1)}rs}+\frac{1}{2\left( n+1\right)
}\left( h^{^{(0)}kr}D_{a}\Gamma
_{bk}^{^{(1)}s}+h^{^{(0)}ks}D_{a}\Gamma
_{bk}^{^{(1)}r}\right)   \nonumber \\
&=&D_{a}D_{b}X^{rs}+\frac{1}{2\left( n+1\right) }\left(
h^{^{(0)}kr}D_{a}\Gamma _{bk}^{^{(1)}s}+h^{^{(0)}ks}D_{a}\Gamma
_{bk}^{^{(1)}r}-D_{a}D_{b}h^{^{(1)}rs}\right) \nonumber \\
&=& D_{a}D_{b}X^{rs}  \label{f9a}
\end{eqnarray}%
because the second term is identically zero. To see this note
that
\begin{eqnarray}
&&h^{^{(0)}kr}D_{a}\Gamma _{bk}^{^{(1)}s}+h^{^{(0)}ks}D_{a}\Gamma
_{bk}^{^{(1)}r}-D_{a}D_{b}h^{^{(1)}rs}  \nonumber \\
&=&D_{a}D_{b}h^{^{(1)}rs}+\frac{1}{2}D_{a}\left(
D^{r}h_{b}^{^{(1)}s}-D^{s}h_{b}^{^{(1)}r}\right)
+\frac{1}{2}D_{a}\left(
D^{s}h_{b}^{^{(1)}r}-D^{r}h_{b}^{^{(1)}s}\right)
-D_{a}D_{b}h^{^{(1)}rs}
\nonumber \\
&=&0.  \label{f9b}
\end{eqnarray}%

There will be three kinds of second-derivative terms that occur.
These are
\begin{eqnarray}
D_{a}D_{b}X^{ab} &=&\frac{1}{n\left( n+1\right) } \times \nn \\ && \left[ \frac{1}{n}%
D_{a}D_{b}\left( D^{a}D^{b}\sigma \right) -\frac{1}{2}D_{a}D_{b}h^{^{(1)}ab}+%
\frac{1}{2}D^{2}\left( \left( n+2\right) \sigma +\frac{n}{2}%
h^{^{(1)}}\right) \right],  \label{f10a} \\
D^{2}X^{ab} &=&\frac{1}{n\left( n+1\right) } \times \nn \\ && \left[
\frac{1}{n}D^{2}\left(
D^{a}D^{b}\sigma \right) -\frac{1}{2}D^{2}h^{^{(1)}ab}+\frac{1}{2}%
D^{2}\left( \left( n+2\right) \sigma +\frac{n}{2}h^{^{(1)}}\right)
h^{^{(0)}ab}\right],  \label{f10b} \\
D_{k}\left( D^{a}X^{kb}\right) &=&\frac{1}{n\left(
n+1\right) }\times \nn \\ &&\left[ \frac{1}{n}D_{k}D^{a}\left( D^{k}D^{b}\sigma \right) -%
\frac{1}{2}D_{k}D^{a}h^{^{(1)}kb}+\frac{1}{2}D^{a}D^{b}\left( \left(
n+2\right) \sigma +\frac{n}{2}h^{^{(1)}}\right) \right]. \nn \\
\label{f10c}
\end{eqnarray}%
To simplify further we need to commute the derivatives. We have,
using (\ref{Riem0}),
\begin{eqnarray}
D_{a}D_{b}\left( D^{a}D^{b}\sigma \right) &=&D_{a}D^{2}\left(
D^{a}\sigma \right) =D_{a}\left[ D^{2},D^{a}\right] \sigma
+D^{2}D^{2}\sigma  \nonumber
\\
&=&-D_{a}\left( \mathcal{R}^{(0)}{}^{ac}{}_{c}{}^{d}D_{d}\sigma
\right) +D^{2}D^{2}\sigma  \nonumber \\
&=&\left( n+1\right) D^{2}\sigma +D^{2}D^{2}\sigma,  \label{f11a} \\
D^{2}\left( D^{a}D^{b}\sigma \right) &=&\left[ D^{2},D^{a}\right]
D^{b}\sigma +D^{a}\left[ D^{2},D^{b}\right] \sigma +D^{a}D^{b}\left(
D^{2}\sigma \right)  \nonumber \\
&=&-\mathcal{R}^{(0)e}{}_{c}{}^{ca}D_{e}D^{b}\sigma -\mathcal{R}^{(0)eb}{}_{c}{}^{a} D^{c}D_{e}\sigma -D^{c}\left( \mathcal{R}^{(0)eb}{}_{c}{}^{a} D_{e}\sigma \right)\nn \\  & & -D^{a}\left( \mathcal{R%
}^{(0)e}{}_{c}{}^{cb}D_{e}\sigma \right) +D^{a}D^{b}\left(
D^{2}\sigma \right)  \nonumber \\
&=&2\left( n+1\right) D^{a}D^{b}\sigma - 2\mathcal{R}_{\phantom{abcd}%
}^{^{(0)}ebca}D_{c}D_{e}\sigma +D^{a}D^{b}\left( D^{2}\sigma
\right),
\label{f11b} \\
D_{k}D^{a}\left( D^{k}D^{b}\sigma \right) &=&\left[
D_{k},D^{a}\right] \left( D^{k}D^{b}\sigma \right) +D^{a}\left[
D^{2},D^{b}\right] \sigma
+D^{a}D^{b}\left( D^{2}\sigma \right)  \nonumber \\
&=&-\mathcal{R}^{(0)ec}{}_{c}{}^{a}D_{e}D^{b}\sigma -\mathcal{R%
}_{\phantom{abcd}}^{^{(0)}ebca}D_{c}D_{e}\sigma -D^{a}\left(
\mathcal{R}^{(0)ec}{}_{c}{}^{b}D_{e}\sigma \right) +D^{a}D^{b}\left(
D^{2}\sigma \right)  \nonumber \\
&=&2\left( n+1\right) D^{a}D^{b}\sigma +D^{a}D^{b}\left( D^{2}\sigma
\right) -\mathcal{R}_{\phantom{abcd}}^{^{(0)}ebca}D_{c}D_{e}\sigma.
\label{f11c}
\end{eqnarray}
Putting this all together we obtain%
\begin{eqnarray}
&&\left[ \mathcal{D}^{2}\tilde{Q}^{ab}+h^{ab}\mathcal{D}_{k}\mathcal{D}_{l}%
\tilde{Q}^{kl}-\mathcal{D}_{k}\left( \mathcal{D}^{a}\tilde{Q}^{kb}+\mathcal{D%
}^{b}\tilde{Q}^{ka}\right) \right] ^{^{(1)}}  \nonumber \\
&=&D^{2}X^{ab}+h^{^{(0)}ab}D_{r}D_{s}X^{rs}-\mathcal{D}_{k}\left( \mathcal{D}%
^{a}X^{kb}+\mathcal{D}^{b}X^{ka}\right)   \nonumber \\
&=&\frac{1}{n\left( n+1\right) }\left[ \frac{1}{n}\Bigg{(} \big{(} (
n+1) D^{2}\sigma +D^{2}D^{2}\sigma \big{)} h^{^{(0)}ab}+2(
n+1) D^{a}D^{b}\sigma \nn \right. \\ & & - 2\mathcal{R}_{\phantom{abcd}%
}^{^{(0)}ebca}D_{c}D_{e}\sigma +D^{a}D^{b}\left( D^{2}\sigma \right)
\Bigg{)}
   \nonumber \\
&&-\frac{2}{n}\left( 2\left( n+1\right) D^{a}D^{b}\sigma
+D^{a}D^{b}\left(
D^{2}\sigma \right) -\mathcal{R}_{\phantom{abcd}}^{^{(0)}ebca}D_{c}D_{e}%
\sigma \right)   \nonumber \\
&&-\frac{1}{2}\left( D^{2}h^{^{(1)}ab}+h^{^{(0)}ab}D_{r}D_{s}h^{^{(1)}rs}-%
\mathcal{D}_{k}\left( \mathcal{D}^{a}h^{^{(1)}kb}+\mathcal{D}%
^{b}h^{^{(1)}ka}\right) \right)   \nonumber \\
&&\left. +D^{2}\left( \left( n+2\right) \sigma
+\frac{n}{2}h^{^{(1)}}\right)
h^{^{(0)}ab}-D^{a}D^{b}\left( \left( n+2\right) \sigma +\frac{n}{2}%
h^{^{(1)}}\right) \right]   \nonumber \\
&=&\frac{1}{n\left( n+1\right) }\left[ -\frac{n+2}{n}D^{a}D^{b}\sigma +\frac{%
1}{2}\left( n+2\right) h^{^{(1)}ab}-\left( \left( n+2\right) \sigma +\frac{n%
}{2}h^{^{(1)}}\right) h^{^{(0)}ab}\right]   \label{f12}
\end{eqnarray}
where (\ref{Riem0}) was used.

From (\ref{f12}) and (\ref{f4}) we see that \bea
\left[\Delta^{ab}\right]^{(1)} &=& \Bigg{[} \hat{K}^{ab}-2\tilde{L}^{cd}\left( \hat{K}%
_{cd}\hat{K}^{ab}-\hat{K}_{c}^{\phantom{c}a}\hat{K}_{d}^{\phantom{c}b}\right) +%
\mathcal{D}^{2}\tilde{L}^{ab}+h^{ab}\mathcal{D}_{k}\mathcal{D}_{l}\tilde{L}%
^{kl}  \nn \\ & & -\mathcal{D}_{k}\left( \mathcal{D}^{a}\tilde{L}^{kb}+\mathcal{D}^{b}%
\tilde{L}^{ka}\right) \Bigg{]} ^{^{(1)}} \nn  \\ &=& 0.  \label{f13}
\eea
In other words, all terms cancel at first order. This is
actually a more general result than we have claimed up to this
point, because no assumptions were made about the the Riemann tensor
for the metric $h_{ab}^{_{(0)}}$. The boundary conditions require
$R_{ab}^{_{(0)}} = (d-2)\,h_{ab}^{(0)}$ but leave the other degrees
of freedom in the Riemann tensor unspecified. Our derivation applies
for any metric $h_{ab}^{_{(0)}}$ with the appropriate Ricci tensor,
even if it has a non-vanishing Weyl curvature.

For the second order terms, using (\ref{l5c}) we have
\begin{eqnarray}
\left[ \mathcal{D}_{a}\mathcal{D}_{b}\tilde{Q}^{rs}\right] ^{^{(2)}}
&=&D_{a}D_{b}\tilde{Q}{}^{^{(2)}rs}+\frac{1}{2\left( n+1\right)
}\left( h^{^{(0)}kr}D_{a}\Gamma
_{bk}^{^{(2)}s}+h^{^{(0)}ks}D_{a}\Gamma
_{bk}^{^{(2)}r}\right)   \nonumber \\
&=&\frac{1}{2\left( n+1\right) }\left( h^{^{(0)}kr}D_{a}\Gamma
_{bk}^{^{(2)}s}+h^{^{(0)}ks}D_{a}\Gamma
_{bk}^{^{(2)}r}-D_{a}D_{b}h^{^{(2)}rs}\right)   \nonumber \\
&=&0,  \label{f14}
\end{eqnarray}%
where the last line follows because the calculation is identical to
that given in eq (\ref{f9b}) above. Consequently (\ref{f5}) and
(\ref{f14}) yield \be \left[\Delta^{ab}\right]^{(2)} = 0. \ee

\subsection{The $d>4$ Boundary Stress Energy}
\label{section:a4}

The boundary stress tensor is given by
\be T^{ab} = \frac{1}{8\pi G }\,\left(\pi
^{ab}-\hat{\pi}^{ab}\right) \ee
up to at least second order in the asymptotic expansion. We now compute the explicit form of $T^{ab}$. Using
\begin{equation}
\pi ^{ab}-\hat{\pi}^{ab} =\frac{1}{\rho ^{3}}\left(h^{ab}\,(P - \hat{P}) -
P^{ab} + \hat{P}^{ab}\right)   \label{f16}
\end{equation}%
we find
\begin{equation}
\left[ P^{ab}-h^{ab}P+h^{ab}\hat{P}-\hat{P}^{ab}\right]
^{{(0)}} = h^{{(0)}ab}-\left( n+2\right)
h^{{(0)}ab}-h^{{(0)}ab}+\left( n+2\right) h^{{(0)}ab}=0,
\label{f17}
\end{equation}%
\begin{eqnarray}
\left[ P^{ab}-h^{ab}P+h^{ab}\hat{P}-\hat{P}^{ab}\right] ^{{(1)}}
&=&P^{{(1)}ab}-\hat{P}^{{(1)}ab}+h^{{(0)}ab}\left( \hat{P}%
^{{(1)}}-P^{{(1)}}\right)   \nonumber \\
&=& - \sigma \, h^{{(0)}ab} - \frac{n-2}{2}\,h^{{(1)}ab}-\frac{1}{n} \, %
D^{a}D^{b}\sigma - \frac{1}{2} \, h^{{(1)}ab}  \nonumber \\
&& + h^{{(0)}ab}\,\left( \left( n+2\right) \sigma +\frac{n-2}{2}\,%
h^{{(1)}}+ \left( \frac{1}{n}\,D^{2}\sigma
 + \frac{1}{2} \, h^{{(1)}}\right)
\right)   \nonumber \\
& = & - \sigma \, h^{{(0)}ab} - \frac{n-1}{2}\,h^{{(1)}ab} - \frac{1}{n}\,%
D^{a}D^{b} \sigma,   \label{f18}
\end{eqnarray}%
and
\be \left[ P^{ab}-h^{ab}P+h^{ab}\hat{P}-\hat{P}^{ab}\right]
^{{(2)}}
=P^{{(2)}ab}- \hat{P}^{{(2)}ab} + h^{{(0)}ab} \, \left( \hat{P}%
^{{(2)}} - P^{{(2)}} \right) = - \frac{n+1}{2}\,h^{{(2)}ab}.
\label{f19} \ee Putting this together yields \be
T^{ab} =  \frac{1}{\rho^{\,d}}\,\frac{1}{8\pi G }\,\left(\sigma \, h^{{(0)}ab}+\frac{d-4}{2}\,h^{{(1)}ab}+\frac{1}{d-3}\,%
D^{a}D^{b}\sigma\right)  +
\frac{1}{\rho^{\,d+1}}\,\frac{d-2}{16 \pi G }\,h^{{(2)}ab} + \ldots . \ee

\setcounter{equation}{0}
\section{A Collection of Useful Identities}
\label{collection}

By commuting derivatives several times, one can prove a number of identities involving $\s$ and its derivatives.  Below we present a collection of such identities in $d=4$. All of these identities, in one form or another, are used in the main text and in appendices \ref{section:b2} and \ref{section:b3}. For ease of reference we state our conventions again: $\s_{abc} := D_c D_b D_a \s$, $\s_{abcd} := D_d D_c D_b D_a \s$, etc.  Also, recall that the commutator of two covariant derivatives acting on a tensor $t_{ab}$ is
\be
[D_{a}, D_{b}]\,t_{c}{}^{d} = R^{(0)}_{abc}{}^{e}t_{e}{}^{d} + R^{(0)}_{ab}{}^{d}{}_{e}t_{c}{}^{e},
\ee
where
\be
R^{(0)}_{abcd} = h^{(0)}_{ac}\,h^{(0)}_{bd} - h^{(0)}_{bc} \, h^{(0)}_{ad}~. \label{Riem0}
\ee
One can show using the equation of motion for $\s$ that
\bea
\s_{c}{}^{c} &=& -3 \,\s \, , \\
\s^{c}{}_{ca} &=& -3 \,\s_a \, \\
\s_{ac}{}^{c} &=& - \s_a \, , \\
\s_e\s^{men} &=& \s_e \s^{mne} - \s^m \s^n + \s^e \s_e \,h^{(0) \: mn} \, ,  \\
\s_{pmn} \s^{pnm} &=& \s_{pmn}\s^{pmn} - 2 \,\s_m \s^m \, , \\
\s^{pq}\s_{apq} &=& \s^{pq}\s_{pqa} - 3 \,\s \s_a - \s_{a}{}^{p}\s_p \, , \\
\s_{pq(a} \s_{b)}{}^{pq}&=& \s_{pq(a}\s^{pq}{}_{b)} - 3 \,\s_a \s_b - \s^q \s_{q(ab)}   \ , \\
\s_{(a}{}^{mn}\s_{b)mn} &=& \s_{mn(a}\s^{mn}{}_{b)} -2 \, \s^n \s_{n(ab)} - 5\,\s_a \s_b + h^{(0)}_{ab}\,\s_m \s^m  \, , \\
\s_{abc}{}^{c} &=& 6 \,\s \,h^{(0)}_{ab} + 3 \,\s_{ab} \, , \\
\s_{(a}{}^{n}{}_{b)n} &=& 2 \,\s_{ab} + 3 \, \s \,h^{(0)}_{ab} \, ,\\
\s^{pmn}{}_{m} \s_{pn} &=& 2 \, \s_{mn}\s^{mn} - 9 \, \s^2 \, , \\
\s_{p(ab)q}\s^{pq} &=& \s_{pq(ab)}\s^{pq} - 6 \,\s \s_{ab} - h^{(0)}_{ab} \, \s_{pq}\s^{pq} - \s_{ap}\s^{p}{}_{b} \, \\
\s^{pq}\s_{pqab}{}^{b} &=& -30 \,\s \s_a + 9 \,\s^{pq}\s_{pqa} \, , \\
\s^{pq}\s_{pqba}{}^{b} &=& -24 \,\s \s_a + 7 \,\s^{pq}\s_{pqa} \, .
\eea
Finally, recall that for any tensor $t_{ab}$
\be
{\mathcal{ D}}_{a}t_{b}{}^{c} = D_a t_{b}{}^{c}
+ \frac{1}{\rho} \, C^{(1)}{}^{c}_{ae} \, t_{b}{}^{e}
-\frac{1}{\rho} \, C^{(1)}{}^{e}_{ab} \, t_{e}{}^{c}
+ \frac{1}{\rho^2} \, C^{(2)}{}^{c}_{ae} \, t_{b}{}^{e}
- \frac{1}{\rho^2} \, C^{(2)}{}^{e}_{ab} \, t_{e}{}^{c}
+  \ldots \, ,
\ee
where expressions for $C^{(1)}{}^{c}_{ab}$ and $C^{(2)}{}^{c}_{ab}$ are
\begin{eqnarray}
 C^{(1)}{}^{c}_{ab} &=&   - h_b^{(0)\,c} \,\sigma_a -
h^{(0)\,c}_a \, \sigma_b + \sigma^c \, h^{(0)}_{ab},  \label{c1}\\
C^{(2)}{}^{c}_{ab} &=& - 2 \,\s \,\left( h^{(0)\,c}_a \,\s_b + h^{(0)\,c}_b \,\s_a -
h^{(0)}_{ab} \,\s^c \right) + \frac{1}{2} \,\left( D_a h^{(2)\,c}_b + D_b
h^{(2)\,c}_a - D^c h^{(2)}_{ab}\right) \label{c2}.
\end{eqnarray}
These expressions were also given in equations (D.12) and (D.13) of \cite{MMV}.


\end{document}